# GUICop: Approach and Toolset for Specification-based GUI Testing

Dalal Hammoud, Fadi A. Zaraket, and Wes Masri*
American University of Beirut
Electrical and Computer Engineering Department
Beirut, Lebanon 1107 2020
e-mails: {*dsh07*, *fz11*, *wm13**}*@aub.edu.lb*

**Abstract**

Oracles used for testing *graphical user interface* (GUI) programs are required to take into consideration complicating factors such as variations in screen resolution or color scheme when comparing observed GUI elements to expected GUI elements. Researchers proposed *fuzzy* comparison rules and computationally expensive image processing techniques to tame the comparison process since otherwise the naïve matching comparison would be too constraining and consequently impractical.

Alternatively, this paper proposes *GUICop*, a novel approach with a supporting toolset that takes (1) a GUI program and (2) user-defined GUI specifications characterizing the *rendering behavior* of the GUI elements, and checks whether the execution traces of the program satisfy the specifications.

*GUICop* comprises the following: 1) a *GUI Specification Language*; 2) a *Driver*; 3) *Instrumented GUI Libraries*; 4) a *Solver*; and 5) a *Code Weaver*. The user defines the specifications of the subject GUI program using the *GUI Specification Language*. The *Driver* traverses the GUI structure of the program and generates events that drive its execution. The *Instrumented GUI Libraries* capture the GUI execution trace, i.e., information about the positions and visibility of the GUI elements. And the *Solver*, enabled by code injected by the *Code Weaver*, checks whether the traces satisfy the specifications.

*GUICop* was successfully evaluated using four open source GUI applications that included eight defects, namely, *Jajuk*, *Gason*, *JEdit*, and *TerpPaint*.

## 1. Introduction

Testing of *graphical user interface* (GUI) programs entails several challenges that have no counterpart when testing text-based command line programs. Most importantly, practical oracles that accurately check whether the observed behavior in GUI execution traces satisfies the expected behavior are hard to construct. In particular, the *rendering behavior* of GUI components; i.e., their appearance and relative positioning, depends on variant non-functional display parameters such as screen resolution, color scheme, line style, thickness and transparency attributes. This necessitates researchers to suggest the use of sophisticated comparison methods such as computationally expensive image processing techniques [4][31][8] and fuzzy comparison rules [23] in oracles. Otherwise absolute comparison would be too constraining and consequently impractical. It should be noted that many researchers opted to circumvent this issue by relying on the *null-oracle*, which considers a program to have failed if it never terminates or terminates abnormally [7].

Other complications do exist. For example, GUI programs have several entry points enabled by an obscure system or library event loop whereas a text-based program has one entry point. Each GUI element accepts sequences of inputs of various types and from various devices as opposed to a fixed number of parameters with predefined types in text-based programs.

Alternatively, this paper proposes *GUICop*, a novel approach with a supporting toolset that takes a GUI program and user-defined GUI specifications that characterize the rendering behavior of GUI elements, and checks whether the execution traces of the program satisfy the specifications. In other words, the specifications act as configurable test oracles. They aim at describing how GUI elements are meant to be displayed in terms of their layout, relative positioning, and visibility.

The *GUICop* supporting toolset includes: 1) a *GUI Specification Language*; 2) a GUI test *Driver*; 3) *Instrumented GUI Libraries*; 4) a *Solver*; and 5) a *Code Weaver*. The user defines the specifications of the subject program using the *GUI Specification Language* whose atomic alphabet consists of basic geometric objects describing GUI components, and positional operators that express relative object positions. We also provide a library of commonly used GUI elements specified in the *GUI Specification Language*. These are used in a hierarchical manner to specify more complex GUI elements and behaviors. Based on the user-defined specifications, the *Code Weaver* injects code at various locations in the subject program. The injected code starts and stops the instrumented output, and invokes the *Solver*. This enables the *Solver* to monitor the execution traces and check the specifications at appropriate locations and times. The GUI test *Driver* traverses the GUI structure of the program in order to generate events that drive its execution [12]. During the execution of the program: 1) the *Instrumented GUI Libraries* capture the GUI execution trace [29] that comprises information about the relative positions taken by the displayed GUI components and the relative times when the GUI events were triggered; and 2) the *Solver* checks whether the captured traces satisfy the user-defined specifications.

A sizable body of work on GUI testing was conducted in the past two decades. The most notable was the work of Memon et al. [11][12][13][14][15][16][32][33] which focused on test case generation, fault-detection, coverage, and regression testing, and the earlier work of Lee White [34] which tackled regression testing of GUI systems. In regard to specification-based GUI testing, the closely related *Pattern-Based GUI Testing* (PBGT) of Paiva et al. stands out [18][19][20][21][22]. PBGT mainly aims at modeling GUI functional requirements; the authors comparatively discuss *GUICop* and PBGT in Section 3. Abbot (*abbot.sourceforge.net*) is an existing specification-based GUI testing framework that is an extension of JUnit. It supports writing specifications for programmable Java GUI components but stops short of enabling the user to specify general layout and component interactions. For example, a component may match its programmable specification, even if it was partially hidden by another component on the screen. Other JUnit extensions that enable the user to write assertions also suffer from that problem, namely, JFCUnit (*jfcunit.sourceforge.net*), Pounder (*pounder.sourceforge.net*), Marathon (*marathontesting.com*), SWTBot (*swtbot.org*), UISpec4J (*www.uispec4j.org*), and Jemmy (*jemmy.java.net*). This paper is an extension of the authors' work presented in [36] and described in the related work section.

Many of the existing tools leverage the GUI hierarchical tree structure where nodes are GUI elements such as frames, text boxes, and push buttons, and edges represent parenthood relations. Generally, these tools require naming all concerned GUI elements and then take the following steps: They (1) find a GUI component of interest starting from the root of the GUI tree based on the *programmable name* of the component; (2) exercise a relevant event on the component; and (3) check the status of the GUI tree following the event using JUnit assertions. However, unlike *GUICop*, the above steps suffer from the following problems:

- *Programmable component names are not always known.* Developers do not always name their GUI components. And even if they do, the names are not necessarily known by testers. Also GUI components could be automatically generated, for example, a scroll bar in an edit box gets instantiated when the length of the text exceeds the width of the edit box.
- *GUI trees are not adequately expressive.* GUI trees capture parenthood information amongst visible components, i.e., they express positional *containment* only and fall short of expressing other positional and timing relations. For example, a YES/NO dialog box may contain the title bar, the message label box, and the YES/NO push button components. While it is easy to express and check such containment relations using a GUI tree, it is not possible to express and check the layout of the components, e.g. YES *is to the left of* NO.

In practice, the authors envision *GUICop* to be primarily deployed as a configurable oracle that monitors the rendering behavior of a GUI program with respect to test cases and user-defined specifications. It is also ideal for making sure that rendering scenarios of interest are correctly designed and covered. Those scenarios could originate from design use cases or from fixes of bugs that should not resurrect [1]. The description of a use case or a bug fix could be readily translated into a GUI specification; and in future work the researchers intend to automate that process [1][28].

This paper makes the following contributions:

- A new GUI specification-based testing approach and supporting toolset that circumvents non-functional discrepancies. Those discrepancies typically hinder the task of reusing test suites, such as changes in screen resolution.
- A novel specification language that enables capturing information about the layout and appearance of GUI components.
- A solver that monitors a GUI execution via instrumentation and code weaving, in order to check whether the GUI application satisfies its user-defined specifications. In its current implementation, *GUICop* supports specifications that capture the expected rendering behavior of GUI components; noting that support for temporal behavior of GUI actions could easily be provided.

- An extensible library of specifications of common GUI components, which allows for the reuse of specifications.
- The most up-to-date implementation of *GUICop* supports the Java Swing graphics library, noting that our prior implementation [36] supported the C++ Qt library.

The remainder of this paper is organized as follows. Section 2 motivates the work. Section 3 discusses related work. *GUICop* and its components are described in Section 4. The conducted case studies and usability experiments are presented in Section 5. The threats to validity of *GUICop* approach are discussed in Section 6. Finally, Section 7 concludes and discusses future work.

## 2. Motivating Example

This section presents a motivating example showing the advantages of *GUICop* over the common approach of programmatically checking for GUI correctness. Given an edit-box and the associated requirement: "*When the text entered by the user exceeds the width of the edit-box, a horizontal scrollbar should appear*." This requirement could be checked programmatically using the code shown in Figure 1.

At $s_1$, the code specifies the edit-box under test using its explicit name. The condition for when a horizontal scrollbar needs to appear is evaluated at $s_2$. The check for

```
void testDriver() {
s1: EditBox b = new EditBox("MyEditbox");
s2: if (b && (b.text.length * b.font.charwidth ≥ b.width)
s3:     assert(containsHSB(b));
}
boolean containsHSB(GUIComponent guiComponent ) {
  if (!guiComponent) return false;
  if (type(guiComponent) is HScrollbar) return true;
  GUIComponent child = guiComponent.firstChild();
  while(child.isValid()) {
    if (containsHSB(child) ) return true;
    child = guiComponent.nextChild() ;}
  return false;
}
```

**Figure 1 –** Checking programmatically for the scroll bar

whether the requirement is satisfied is done at $s_3$ by calling the containsHSB() method which recursively traverses the GUI tree rooted at the edit-box in order to check whether it contains a horizontal scrollbar. Consider the following problematic points in this code:

1. What if the name of the edit box under test is not known?
2. Writing this code requires that the tester has enough expertise on how the GUI components are represented in the GUI tree.
3. This code needs to be tailored/ported for each supported GUI API such as Qt, MFC, and Swing.
4. What if the GUI tree contains the scrollbar but it is actually not visible on the screen?

Noting that Textrect, Editbox, and HScrollbar are GUI components supported within *GUICop*, the alternative specification would be:

```
EditboxOverflow = {
   variables {
      Textrect t1;
      Editbox eb;
      HScrollbar hb;
   }
   properties {
      X = eb.x;
      Y = eb.y;
      WIDTH = eb.width;
      HEIGHT = eb.height;
   }
   constraints {
     (eb contains t1);
     (t1.width > eb.width) implies ((eb contains hb) and (t1 above hb));
   }
}
```

Focusing on the construct “constraints” above, the specification asserts the following: 1) edit-box eb contains text area t1; and 2) if the width of t1 exceeds the width of eb, then eb contains horizontal scrollbar hb and t1 is rendered above hb.

The following are highlights concerning the EditboxOverflow specification:

1. No explicit component name is needed.
2. The tester does not need to know that a GUI tree even exists.
3. Due to the high level of abstraction at which the specification is written, the specification is portable across machines, displays, and GUI libraries.

4. The issue of component visibility is implicitly taken care of by the instrumented GUI libraries within *GUICop*.

## 3. Related Work

This paper extends the initial work of the authors presented in [36] by the following: 1) significantly improving the solver algorithm to reduce the number of matching objects per AST tree node; 2) implementing the *Code Weaver*; 3) implementing the *Driver*; 4) conducting real life case studies; 5) extending the expressiveness of the *GUICop Specification Language* with new constructs and new library elements; 6) providing a comprehensive description of the *GUICop* toolset; and 7) extending instrumentation support for the *Swing* library so that it produces GUI event traces in addition to the existing C++ Qt library support in [36].

Numerous existing GUI testing tools require the user to manually write unit tests to validate the behavior of the GUI application in order to automate the test execution, such as Abbot, Pounder, JFCUnit, SWTBot and UISpec4J. Other techniques would capture the user sessions and replay them later without having the tester writing unit tests, such as HP WinRunner [6] and jRapture [29]. Using *GUICop* the specifications are defined separately without intertwining them with the unit tests. Other tools like Sikuli for instance, allow testers to take a screenshot of a GUI element and query a help system using the screenshot instead of the name of the component [4][31]. For example, a tester can write the following script: "click(>); assertExist(||); assertNotExist(>);". This script states that when the play button is pressed, it should automatically be replaced by a pause button. Idioms '>' and '||' refer to real snapshots in the Sikuli environment. The main issue in Sikuli is that it's highly dependent on the images in the application, and on the underlying image processing techniques. *GUICop* on the other hand operates at a higher level of abstraction; user specifications make use of predefined GUI objects provided by *GUICop* specifications libraries. Therefore, *GUICop* test oracles do not need to be updated as often. For example, if the play pushbutton changed in the GUI library supporting an application, the corresponding object in the *GUICop* specification library should be changed as well to reflect the new implementation; however; the high level specification need not be changed. Within Sikuli, the user needs to: 1) replace the image of the pushbutton, and

2) make sure that the object identification image processing techniques identify the corresponding objects.

Another technique developed by Memon et al., GUI Ripping [2][9][12], traverses the application's GUI and extracts its structure and execution behavior in order to automatically generate test cases; the *GUICop Driver* is actually implemented based on the work presented in [12]. *GUICop* on the other hand focuses on an instrumented version of the application to capture its behavior. *GUICop,* in comparison with the other tools, is innovative in terms of: 1) the level of abstraction it operates at; 2) the reusability of its specifications; 3) its accuracy as it depends on instrumenting GUI libraries; and 4) the automation potential it provides.

Xie and Memon [35] assessed the impact of the used GUI test oracles on the fault detection effectiveness and cost of a test case. Their main findings were that: 1) fault detection effectiveness diminishes drastically when using "weak" test oracles; 2) checking against a "strong" oracle at the end of test case execution is most cost-beneficial; and 3) frequently checking against "strong" oracles can compensate for not having test cases with long sequences of events. These findings are favorable to our approach, since *GUICop* allows the user to define test oracles exhibiting a wide spectrum of detail (ranging from "very weak" to "very strong").

Researchers have also devised model-based GUI testing approaches, which are closely related to specification-based testing. That is, they generated test cases out of models that characterize GUI programs [13], or they used these models as oracles [3][25]. The models were extracted from the programs via reverse engineering in [12][27][19], and were built manually using specification languages in [25][26] (e.g., VDM [30], Spec#, and PARADIGM [17]). The work most relevant to *GUICop* is *Pattern-Based GUI Testing* (PBGT) [19][20], which promotes the reuse of GUI testing strategies that target GUI functional requirements. In an analogy to design patterns, User interface (UI) patterns and testing patterns represent repeatable solutions to commonly occurring problems in GUI design and testing, respectively. A UI Pattern is a template for how to solve a GUI problem that can be used in slightly different situations. A *UI Test Pattern* [20] provides a configurable test strategy to test an implementation of a given UI Pattern embedded in a GUI program. In PBGT, a UI Test Pattern may be configured to specify how the application should behave, by providing the following: 1) *Test Goals*: identifiers/names of the tests; 2) *V*: the set of variables involved in the test; 3) *A*: the sequence of actions to perform during test execution; 4) *C*: the set of possible checks to perform during test execution.

The authors in [20][18] illustrated the concept of UI Test Patterns using the "Login UI Test Pattern" (among others), which defines a test strategy for the authentication process in GUI applications. They define the "Login UI Test Pattern" as follows:

1) *Test Goals*: {"Valid login", "Invalid login"}. That is, check that the authentication will succeed given a valid username/password, or, check that the authentication will fail given an invalid username/password.
2) *V*: the involved variables are {*username*, *password*}.
3) *A*: the required actions are [provide *username*; provide *password*; press *submit*].
4) *C*: the available checks are {"change to page X", "pop-up error message Y", "stay on same page"}.

Given a GUI that uses the "Login UI Pattern" in which the *username* is labeled as "Email", the *password* as "Password", and the submit button as "LogIn", a tester would configure the "Login UI Test Pattern" as follows:

1) *Test Goal*: "Valid login"
2) *V*: {[ Email, "correctEmail"], [Password,"correctPassword"]}
3) *A*: [provide Email, provide Password, Press LogIn];
4) *C*: {change to page "Welcome"}

Or possibly as below:

1) *Test Goal*: "Invalid login"
2) *V*: {[ Email, "correctEmail"], [Password, "incorrectPassword"]}
3) *A*: [provide Email, provide Password, Press LogIn];
4) *C*: { pop-up error message "Please re-enter your password"}

As described in [17][18][19][20][21][22], PBGT mainly provides capabilities to (1) model GUI functional testing requirements, (2) define reusable testing strategies, and (3) generate test cases out of the specified testing models. While the methods of PBGT require the test suites to satisfy the testing requirements in terms of covering all the test models and consequently testing scenarios specified in them, these methods do not specify how GUI components should be correctly rendered, as it is the case with the *GUICop* specifications. In addition, it appears that PBGT does not provide an automated ability to verify whether a given test case passed or failed, i.e., the tester is required to provide such a verification mechanism. On the other hand, *GUICop* provides such verification ability (via its *Solver*) which takes into consideration the sensitivity and complications of visual artifacts such as variations in screen resolutions, and color schemes. In summary, *GUICop* compares to PBGT as follows: 1) PBGT uses a DSL (Domain Specific Language) to define GUI Models, whereas *GUICop* uses a DSL to define GUI test oracles; 2) PBGT generates test cases from GUI Models, whereas *GUICop* generates them via *GUI Ripping* if needed; 3) test case generation and execution are automated in both; 4) PBGT allows for defining simple oracles, whereas *GUICop*'s oracles could be as complex as the user defines them to be; and 5) *GUICop* provides the ability of automatically checking against the defined oracles, whereas PBGT does not (since related literature does not suggest otherwise [19][20]).

## 4. *GUICop:* Approach and Tool Set

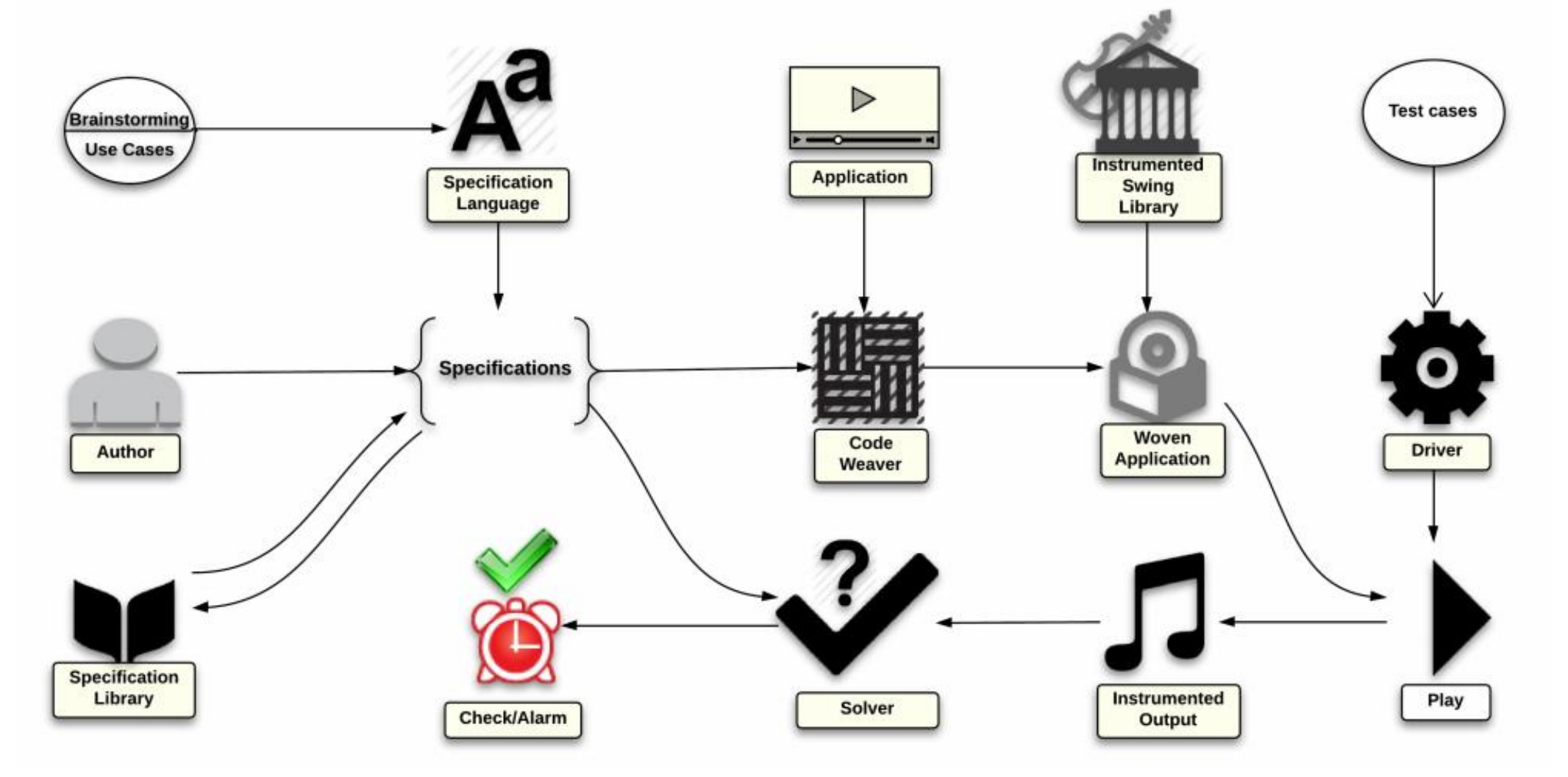

**Figure 2 -** *GUICop* Flow Diagram

Figure 2 provides an overview of the *GUICop* approach. The user writes specifications using the *GUICop Specification Language* to describe the expected rendering of the GUI under test; these user-defined specifications are basically GUI test oracles that characterize the expected output. Note that the creation of a new DSL was necessary due to the required expressiveness of the *GUICop* specifications. A *Specification Library* is maintained in order to support the reuse of specifications of commonly used GUI components that were previously defined by the user. By design, *GUICop* is not meant to modify the SUT's source code; the alternative is provided by the *Code Weaver*, which injects the SUT's code with calls to the *Solver*. The *Instrumented GUI Library* allows for capturing traces of GUI events. The *Driver* executes the SUT in order to generate the traces to be validated by the *Solver*. Specifically, the *Driver* uses GUI Ripping, proposed by Memon et al. [2][9][12], to generate test cases, and the *Solver* checks whether the traces satisfy the specifications to determine whether a given test case passed or failed. Described next, are the main components of *GUICop*, namely, the *Language*, the instrumentation, the *Weaver*, the *Driver*, and the *Solver*.

### 4.1. GUI Specification Language

The *GUICop Specification Language* aims at capturing positional, arithmetic, logic, and relational GUI behaviors. Its design was influenced by brainstorming close to fifty representative GUI components and scenarios which included standalone applications such as *Calculator* and *Music Player*, and GUI components such as *Menu Bar*, *Top Bar*, *Scrollbar* and *Drop Down List*.

Consider a radio button, which in its normal state should appear as an ellipse. But when pressed, it should appear as two ellipses, one inside the other as shown below.

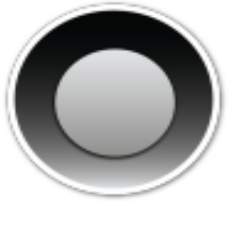

In order to test its behavior for when it is pressed, the user could write the following specification:

```
PushedRadiobutton = {
    variables {
            Ellipse e1, e2;
    }
    properties {
            X = e1.x;
```

```
            Y = e1.y;
            WIDTH = e1.width;
            HEIGHT = e1.height;
        }
        constraints {
            (e1 contains e2);
        }
    }
```

The above asserts that when the button is pushed, ellipses e1 and e2 should appear, the button should be confined to the rectangle bounding e1, and e1 should contain e2. The above example illustrates the following regarding a *GUICop* specification:

1) It comprises the construct "variables" in which the user declares the variables to be used. These variables could be of primitive types, namely, Rectangle, Line, Ellipse, Polygon, Triangle, Text, and Textrect; or they could be of complex types, i.e., of a type previously defined by the user and archived in the *Specification Library*.
2) It comprises the construct "properties" which is a list of name value pairs specifying information such as the location and size of the component under test. The properties include by default the elements of the rectangle that bounds the component, namely, X, Y, WIDTH and HEIGHT.
3) It comprises the construct "constraints" in which the user describes how primitive or complex objects should appear and how they should be positioned with respect to each other. A constraint is an expression involving the declared variables as operands and the supported operators. Noting that several constraints could be defined.

ANTLR (*www.antlr.org*) was used to parse the user-defined specifications in order to generate the corresponding *Abstract Syntax Tree* (AST) that will be processed by the *Solver*. Figure 3 provides the ANTLR grammar where **spec-objects** is the start symbol. As shown, multiple specifications are supported in which the constraints could involve positional, arithmetic, logic, and relational operators. Note that following a symbol with '^' indicates that the given symbol should be the root of the corresponding subtree in the generated AST, shown in Figure 4.

```
spec-objects:  (spec-object)+
  ;

spec-object :  (ID^  '='!  '{'!
                   variables
                    (properties)?
                    constraints
                 '}'!)
  ;
variables  :  ( 'variables'^  '{'!
                    variables-decl*
                 '}'!)
  ;
properties  :  ( 'properties'^  '{'!
                   properties-decl*
                 '}'!)
  ;
constraints :  ( 'constraints'^  '{'!
                    constraints-decl ';'!
                  '}'!)
  ;

variables-decl:  ( ID^  ID (','!  ID)*';'!)
  ;
expression :  member-variable-access (OPERATOR^ member-variable-access)*
  ;
member-variable-access: ID '.'^ PROPERTY | ID '.'^ ID
  ;

properties-decl :  (  (PROPERTY^ | ID | ID'.'^ ID )  '='!  expression ';'! )
  ;

constraints-decl :  '('! constraints-decl  OPERATOR^  constraints-decl ')'!
                          |  '('! NOT constraints-decl ')'!
                          |  member-variable-access
                          |  ID
                          |  INT
                          |  QUOTED-STRING
  ;

OPERATOR:     'leftto'  |'rightto'|'above'|'below'|'contains'|'over'|'smaller'     // positional
          |   'leftaligned'|'rightaligned'|'topaligned'|'bottomaligned'           // positional
          |   'and'    |    'or'    |    'not'   | 'xor' |  'implies'             // logic
          |   '+'      |    '-'     |    '*'     |  '/'                           // arithmetic
          |   '=='     |    '.'     |    '<'     |  '>'  |    '!='                // relational
          |   'equals' |    'concat'                                              // string
          ;
PROPERTY:  'X' | 'x' | 'Y' | 'y' | 'WIDTH' | 'width' | 'HEIGHT' |'height'
;
```

**Figure 3** – Grammar defining the *GUICop Specification Language*

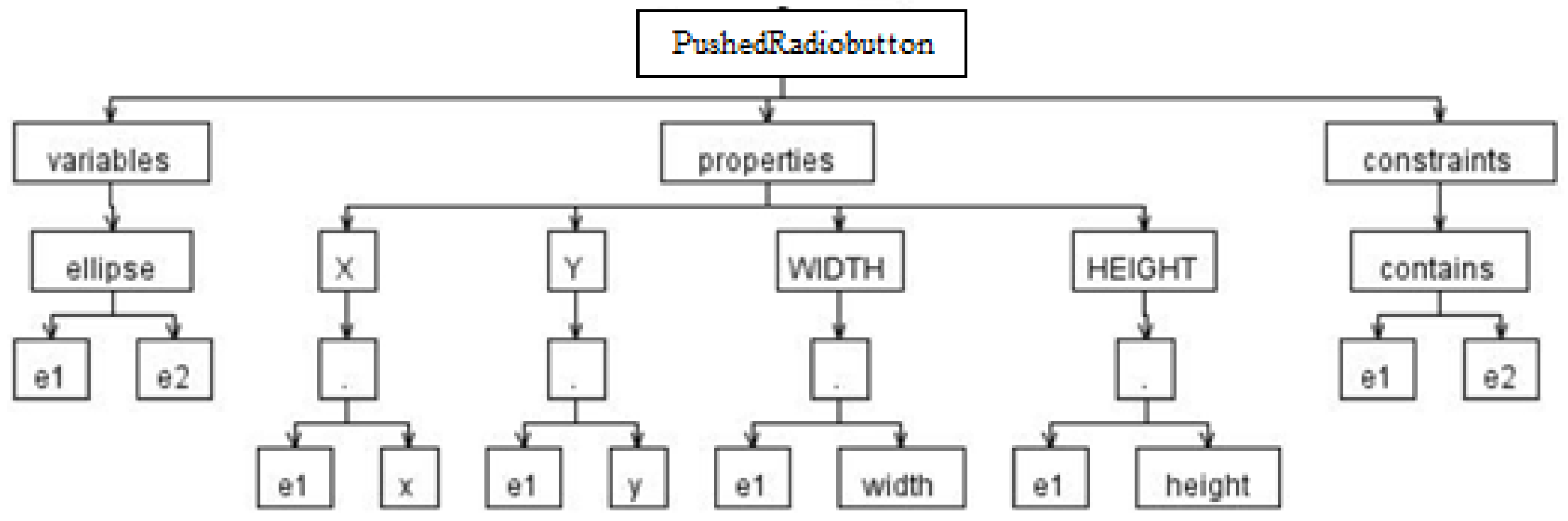

**Figure 4-** Abstract Syntax Tree (AST) of the PushedRadiobutton specification

### 4.2. Instrumenting the GUI Libraries

In order to capture the GUI traces, *GUICop* requires that the underlying GUI library be instrumented. Specifically, code should be injected in functions that draw basic shapes such as Rectangle, Line, Ellipse, Polygon, Triangle, Text, and Textrect. The purpose is to externally save information about the rendered GUI components. A typical trace would take on the following form:

```
rectangle(x, y, w, h);
line(x1, x2, y1, y2);
…
polygon(x1, y1, ..., xn, yn);
triangle(x1, y1, x2, y2, x3, y3);
textrect(x, y, w, h, str);

…
```

The trace data generated by the instrumented graphics library follows a format that adheres to the ANTLR grammar shown in Figure 5. The captured traces are passed to

```
shapes: (shape)*;

shape:
            ('rectangle'^ '(' INT ',' INT ',' INT ',' INT ')' ';')
      |     ('line'^ '(' INT ',' INT ',' INT ',' INT ')' ';')
      |     ('ellipse'^ '(' INT ',' INT ',' INT ',' INT ')' ';')
      |     ('polygon'^ '(' INT ',' INT (',' INT ',' INT)* ')' ';')
      |     ('triangle'^ '(' INT ',' INT ',' INT ',' INT ',' INT ','INT ')' ';')
      |     ('text'^ '(' INT ',' INT ',' STR ')' ';')
      |     ('textrect'^ '(' INT ',' INT ',' INT ',' INT ',' STR ')' ';')
;
```

**Figure 5** – Grammar defining the GUI traces

the *GUICop Solver* to be checked against the specifications, which are provided to the *Solver* in the form of AST's. Note how each shape is associated with a number of attributes such as coordinates, size and text. As an example, consider the simple GUI trace below:

```
rectangle(10, 15, 10, 20);
line(2, 5, 8, 12);
```

The attributes of the rectangle represent the coordinates of the top-left corner, width, and height. The attributes of the line represent the coordinates of the start and end points. Also, the trace provides timing information indicating that the rectangle was rendered before the line; in the future, such information could be leveraged by constraints that use temporal operators.

It should be noted that providing a similar functionality by using the accessibility API of the given GUI library might have been possible. However, it was much easier to implement this functionality via instrumentation to overcome several nitpicky issues that were faced. For example, for optimization purposes graphics libraries may render objects several times on auxiliary devices before actually writing to the screen as in the case of double buffering. Finally, note that the most up-to-date implementation of *GUICop* supports/instruments the Java Swing graphics library.

### 4.3. Code Weaver

The *Code Weaver* is used to deploy the *GUICop* specifications. It uses Aspect Oriented Programming (AOP) to weave calls (using AspectJ) from the SUT to the *Solver* at appropriate code locations. The calls invoke the *Solver* to check whether a given specification (whose identifier is passed as a parameter) has been violated or

```
aspect testscript{
        pointcut drawMenu():execution(void drawMenu());
        after(): drawMenu(){
        guicop.check("MenuItemSeparator");}
}
```

**Figure 6** – Example AspectJ code.

```
title testscript
after drawMenu()
guicop MenuItemSeparator
```

**Figure 7** – Equivalent *GUICop* aspect code.

not.

For example, the code weaver take the input in Figure 7 from a *GuiCop* user. The input includes the name of the specification "MenuItemSeparator", and a code location specified with either a line number or a function name "drawMenu()". The code weaver uses the input with templates of Java aspects to produce the aspect in Figure 6. Then AspectJ is called to inject code into the SUT that calls the GUICop solver. The user can also specify code locations to start, stop, and resume instrumentation output.

### 4.4. Driver

For the purpose of test case generation, the researchers implemented the *GUICop Driver* to support Java Swing applications following the GUI ripping technique proposed by Memon *et al.* [2][9][12]. A GUI structure is generally represented as a forest of trees where each tree is rooted at a top frame or window. Starting at the top level windows (i.e. the windows that are visible when the application first starts), the *Driver* traverses the corresponding trees and triggers the visited GUI components following depth first traversal, as described below.

The *Driver* launches the application under test via Java reflection. The top level visible GUI windows are identified using the java.awt.Window.getWindows() method. A click event is applied on each identified executable top level window and on each of their executable descendants; i.e., windows that derive from the AbstractButton class which provides the doClick() method. In regard to text components such as JTextField and JTextArea, the *Driver* trigger them by typing few characters within following a check of whether they are editable.

Finally, the researchers do not consider the *GUICop Driver* to be a contribution given its similarity to existing work. In addition, it is not truly critical to the operation of *GUICop* since test suites could alternatively be executed manually or using some other automated approach. For example, the case studies presented in Section 5 did not involve the *Driver*.

### 4.5. Solver

The purpose of the *GUICop Solver* is to check whether a GUI trace satisfies the specifications defined by the user. The timing of a given check is dictated by the location at which the *Code Weaver* injected its corresponding call. Multiple checks might be performed since multiple specifications could be defined and each could be weaved at multiple locations. A check is driven by the GUI trace and the AST of the specification at hand, specifically, its constraints subtree. Illustrated next is how the *Solver* performs a check, considering the GUI in Figure 8, the specification in Figure 9, the trace in Figure 10, and the constraints subtree in Figure 11. The subsequent sections describe and discuss the algorithms behind the *Solver*.

### 4.5.1 Illustrative Example

The internal nodes of the constraints subtree represent the operators, and the leaves represent the variables declared in the variables construct of the specification. In this case, the variables are all of primitive types; however, if a variable was of a complex user-defined type, its corresponding leaf node would be replaced by the constraints subtree of its type. Listed below are the steps taken by the *Solver* to perform the satisfiability check in the example:

*Step1.* Each leaf node in Figure 11 is annotated with the objects appearing in the GUI trace (from Figure 10) that match its type. For example, since leaf nodes "r1", "r2", and "r3" represent objects of type Rectangle, each is annotated with all the rectangles captured in the trace, namely, o1, o2, o3, and o4. Similarly, triangles "t1" and "t2" are annotated with the traced triangles o5 and o6. This step provides the initial solution, that is, lists of traced objects that potentially match their respective leaf nodes.

*Step2.* The annotations of *Step1* are augmented with information about the order of appearance of the objects in the variables section of the specification. For example, since t2 was declared second in variables and was initially annotated with o5 and o6, the new annotations become *<j><o5><j><j><j>* and *<j><o6><j><j><j>*. The "*j*" represents a dummy place holder (or joker) to be identified in subsequent steps. The outcome of *Step1* and *Step2* is shown in Figure 11.

*Step3.* The AST is recursively traversed in order to identify and process the subtrees that are rooted at an operator with two leaf nodes. In the example two of such subtrees are identified, the first is rooted at a "contains" operator with r1 and t1 as operands, the second is also rooted at a "contains" operator but with r3 and t2 as operands.

*Step4.* Processing the first subtree involves computing the Cartesian product of the respective solutions of r1 and t1, and for each of the resulting eight pairs checking if the rules associated with the "contains" operator hold. For example, the pair (*<j><j><o3><j><j>*, *<o5><j><j><j><j>*) satisfies the rules for "contains" since (o3.getMostTop() < o5.getMostTop() and o3.getMostLeft() < o5.getMostLeft() and o3.getMostBottom() > o5.getMostBottom() and o3.getMost Right > o5.getMostRight()). Whereas the pair (*<j><j><o1><j><j>*, *<o6><j><j><j><j>*) violates such rules since (o1.getMostTop() > o5.getMostTop()). In all, it is determined that only (*<j><j><o3><j><j>*, *<o5><j><j><j><j>*) and (*<j><j><o2><j><j>*, *<o6><j><j><j><j>*) satisfy the "contains" rules. The second subtree is processed in a similar manner yielding the same results. In order to reflect these results on the AST, the *Solver* collapses the two processed "contains" subtrees and annotates them with *<o5><j><o3><j><j>* and *<o6><j><o2><j><j>* as the updated solutions. The collapse operation is set intersection; e.g., *<o5><j><o3><j><j>* is the set of all 5-tuples where the first and the third elements are *o5* and *o3*, and is the intersection of the sets described by the tuples *<j><j><o3><j><j>* and *<o5><j><j><j><j>*. The *Solver* also computes the new properties (i.e., bounding boxes) for the resulting leaf nodes based on the properties of the components of their respective solutions, namely, o2, o3, o5, and o6. The outcome of *Step3* and *Step4* is shown in Figure 12.

*Step5.* At this point, the subtree to the left of the AST root is the only subtree that is rooted at an operator with two leaf nodes; therefore, the *Solver* will process it next. It is rooted at a "leftto" operator with r2 as a right operand and the leftmost node resulting from *Step4* as a left operand. This subtree is processed similarly to *Step4* except that the rules associated with the "leftto" operator are used instead. The pairs that satisfy the "leftto" rules are found to be ((*<o6><j><o2><j><j>*), *<j><j><j><o4><j>*) and ((*<o6><j><o2><j><j>*), *<j><j><j><o3><j>*), which is reflected in Figure 13 by collapsing the processed subtree and annotating it with *<o6><j><o2><o3><j>* and *<o6><j><o2><o4><j>*. Here also, the *Solver* computes the bounding box for the resulting leaf node based on the bounding boxes of o2, o3, o4 and o6.

*Step6.* Finally, when the *Solver* processes the last remaining subtree, it determines that (*<o6><j><o2><o4><j>*, *<j><o5><j><j><o3>*) is the only pair that satisfies the "leftto" rules. Therefore, it collapses the node and annotates it with *<o6><o5><o2><o4><o3>*, as shown in Figure 14. Given that a solution for the operation associated with the root of the AST was found, the specification is considered to be satisfied. Note how, as expected, the solution did not include *o1*.

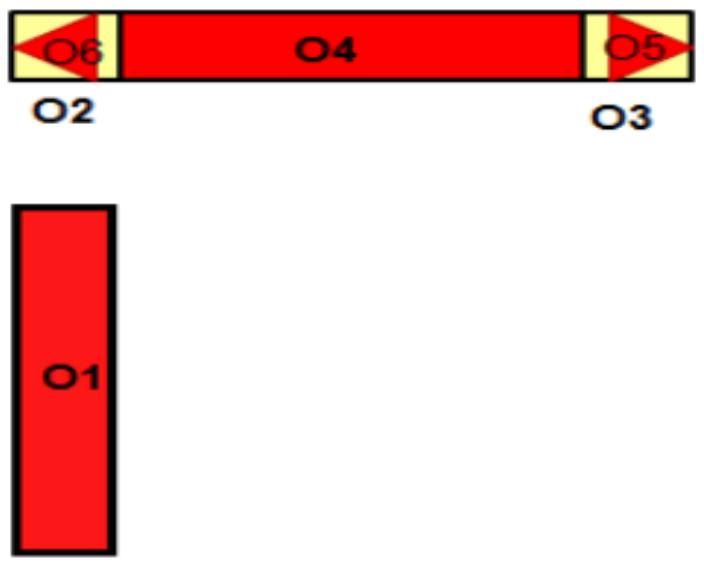

**Figure 8** – GUI under test: horizontal scrollbar

```
HScrollbar = {
        variables {
                Triangle t1, t2;
                Rectangle r1, r2, r3;
        }
        properties {
                X = r1.X;
                Y = r1.Y;
                WIDTH = r1.WIDTH + r2.WIDTH + r3.WIDTH;
                HEIGHT = r1.HEIGHT;
        }
        constraints {
                (((r1 contains t1) leftto r2) leftto (r3 contains t2));
        }
}
```

**Figure 9** – GUICop specification

```
o1: rectangle (10 , 30, 10, 60) ;
o2: rectangle (10 , 10, 10, 10) ;
o3: rectangle (80 , 10, 10, 10) ;
o4: rectangle (20 , 10, 60, 10) ;
o5: triangle (88 ,15 ,80 ,18 ,80 ,12) ;
o6: triangle (12 , 15, 20, 18, 20, 12) ;
```

**Figure 10** – GUI trace

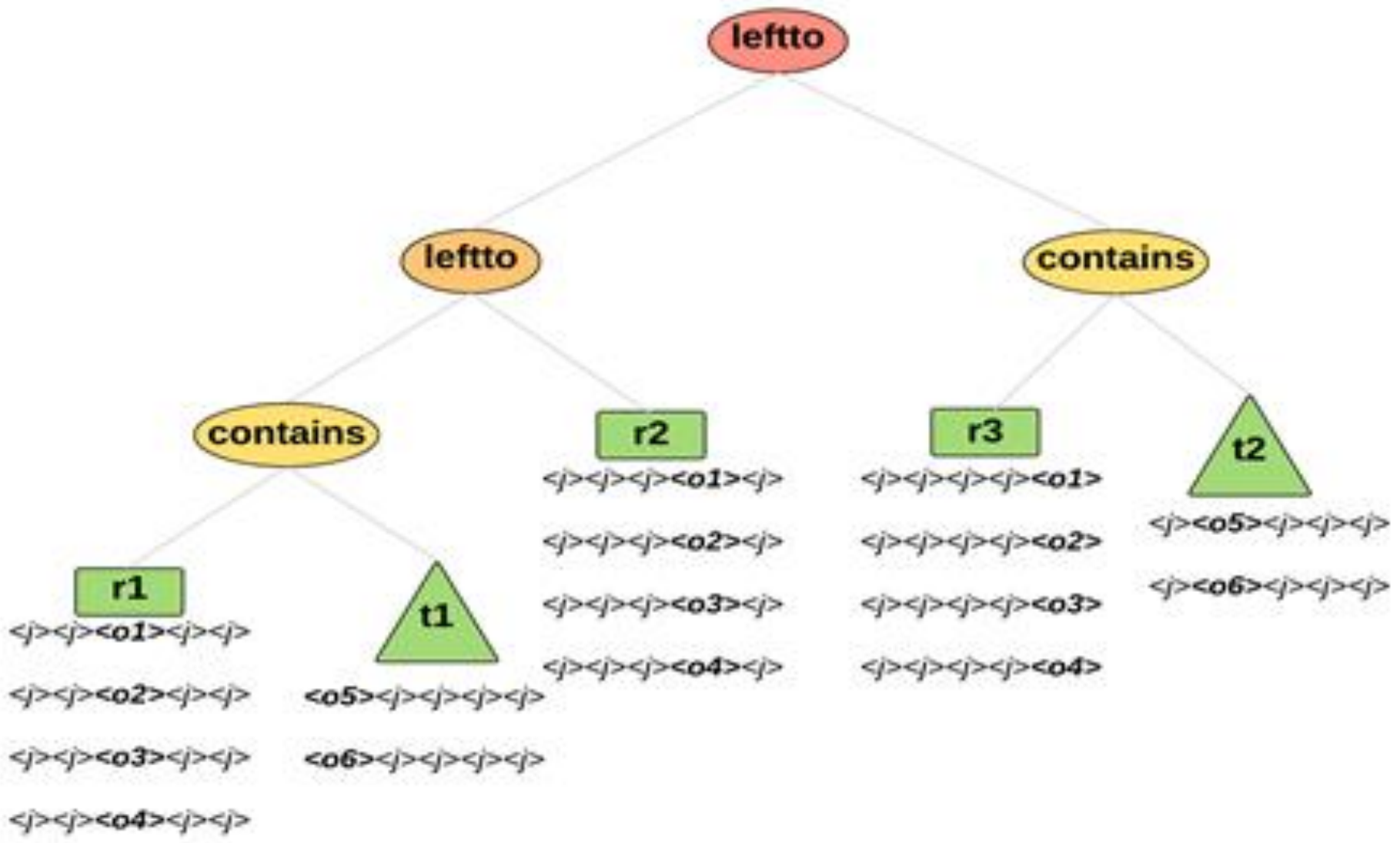

**Figure 11-** “constraints” tree annotated with the outcome of *Step1* and *Step2*

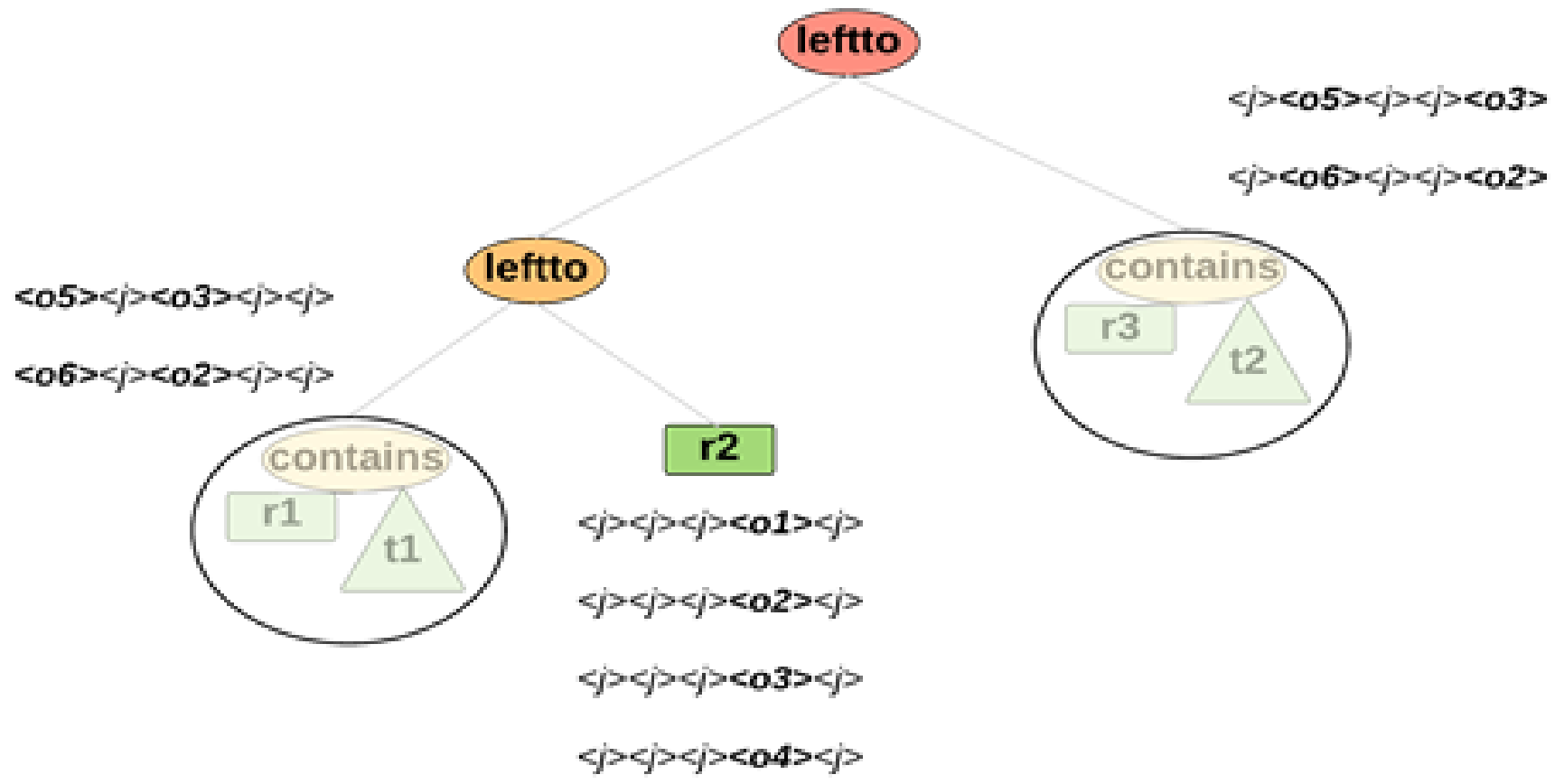

**Figure 12-** *Solver*: *Step3* and *Step4*

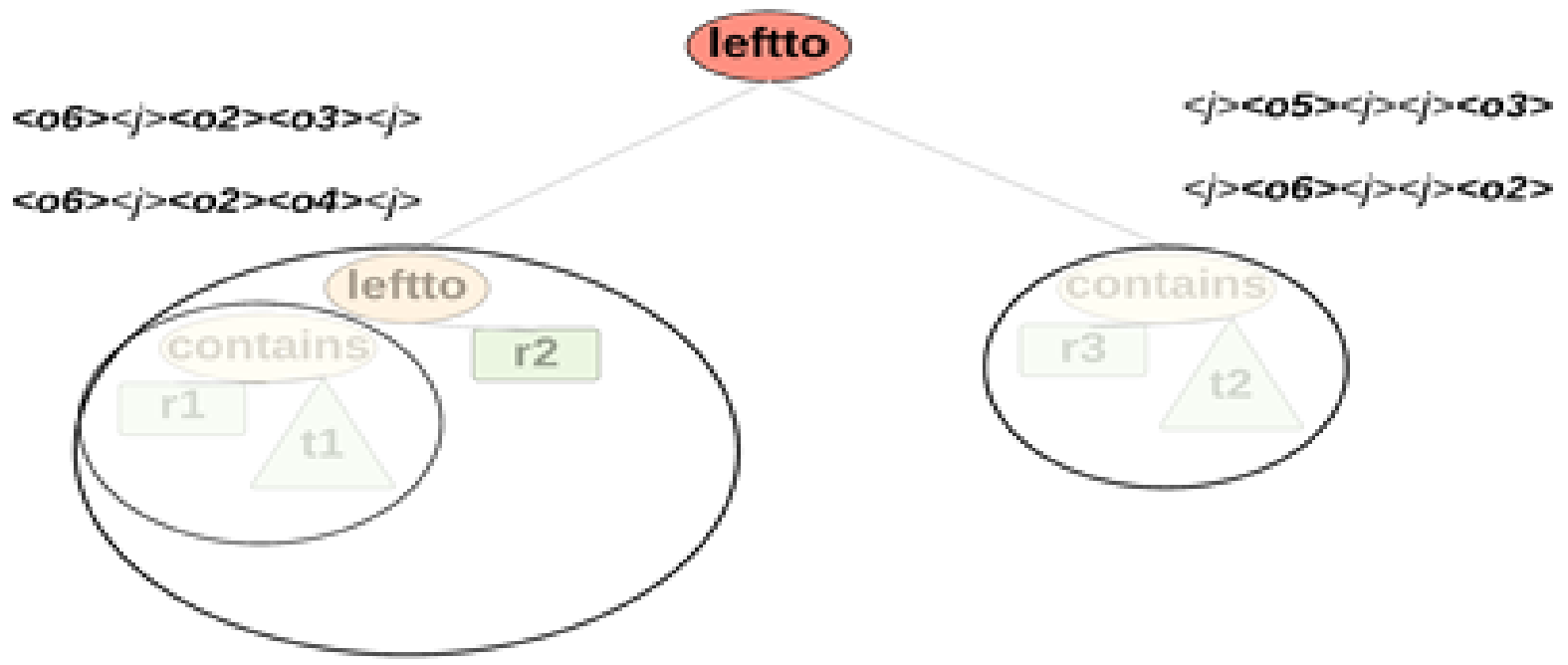

**Figure 13-** *Solver*: *Step5*

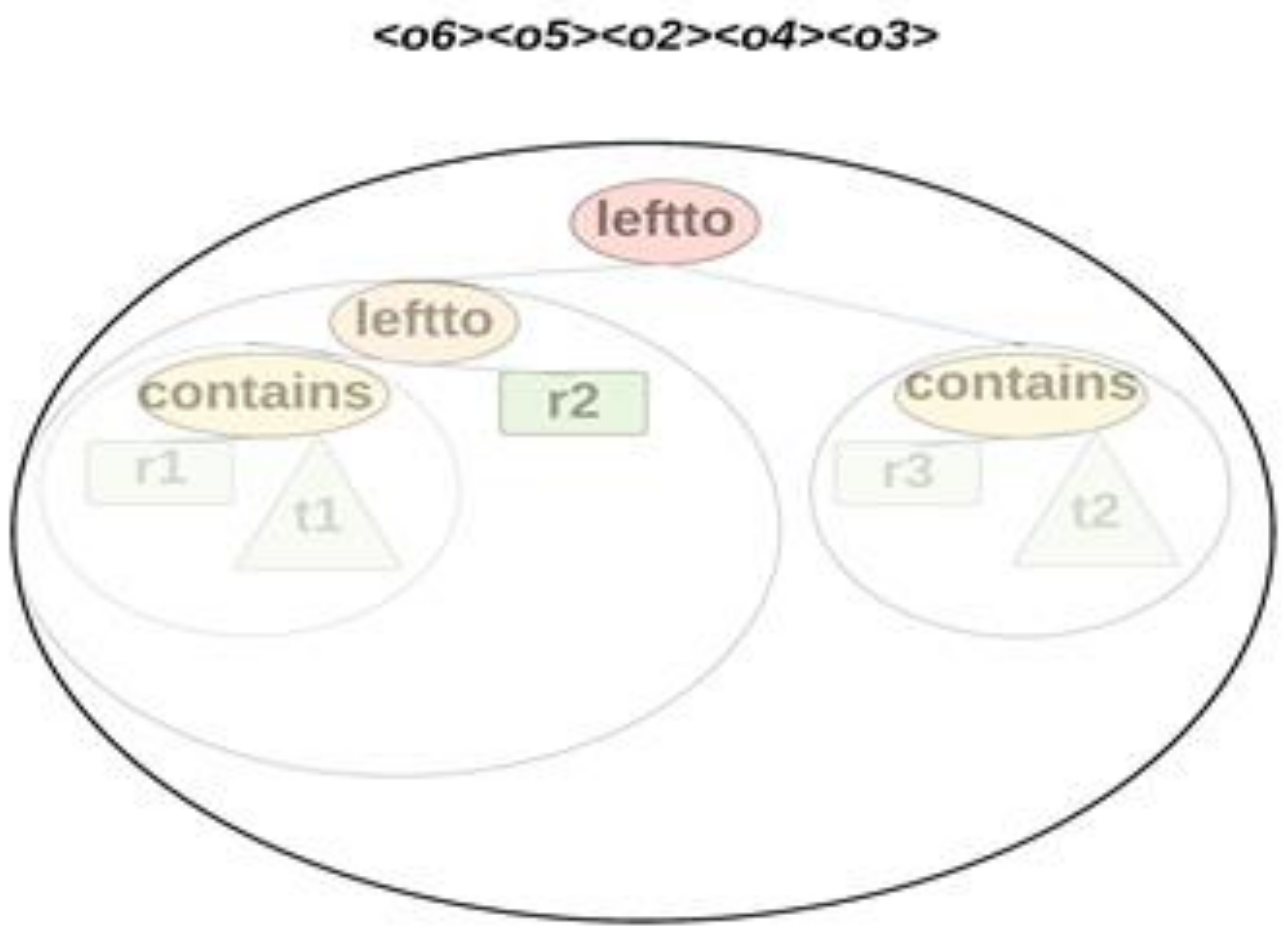

**Figure 14-** *Solver*: *Step6*

***Algorithm-1***

```
Solve (Parent, Left , Right)
1. If isLeaf(Parent) return Parent.Components = components(Parent.type) //base case
2. List1 = Solve(Left, Left.left, Left.right) //recursive call
// handle negation (unary operator)
3. If (Parent.operation == "!") return Parent.components = complement(Left.Components)
// operations AND and IMPLIES can be shortcut in special cases
4. If (List1. isEmpty())
5.     if (Parent.operation == "implies") return Parent.Components = Universe
6.     if (Parent.operation == "and") return Parent.Components = List1
7. end if
8. List2 = Solve(Right, Right.left, Right.right) //recursive call
// logical or is simply set union
9. If (Parent.operation == "or") return Parent.Components = List1 union List2
// apply the operation over all pairs of Cartesian product List1 X List2
10. foreach comp1 in List1 do
11.     foreach comp2 in List2 do
12.         if applyOperation(Parent.operation, comp1, comp2) then
13.             List3.Add( Merge(comp1, comp2) )
14.         end if
15.     end for
16.end for
17.return Parent.Components  = List3
```

***Algorithm-2***

```
Merge (C1 , C2) // C1.Shapes.size == C2.Shapes.size
C3 = new Component
for i = 1; i < C1.Shapes.size do
    comp1 = C1.Shapes(i)
    comp2 = C2.Shapes(i)
    if (comp1.isJoker() ) then   // so add comp2 whether it were a joker or not
        C3.Shapes.AddAtPosition(comp2, i)
    else // comp1 is not a joker
        if ( comp2.isJoker() or comp1 == comp2) then
            C3.Shapes.AddAtPosition( comp1, i)
        else // conflict denoting that intersection is empty
            return null
        end if
    end if
end for
return C3
```

**Figure 15** – Solve and Merge Algorithms

### 4.5.2 Solver Algorithms

*Algorithm-1* in Figure 15 provides pseudocode for the **Solve** algorithm that recursively processes a constraint node in the specification AST tree. The *base case* is when the node is a leaf which corresponds to a primitive shape. Algorithm **Solve** calls and returns the output of function **components** which in turn returns all shapes and objects in the instrumentation output that correspond to a primitive shape type. In case the node corresponded to a unary operator (negation), **Solve** recursively computes the set of components matching *Left* the negated operand. Then it returns its complement. Otherwise, the node is a binary operator. In case the set of components matching *Left* was empty, operations “And” and “Implies” can be directly computed. Operation “Implies” returns the *Universe*; i.e. all possible components signified by a tuple of all joker elements. Operation “And” returns the empty set; i.e. *List1* itself in this case. For the rest of the operations, **Solve** computes the set of components matching the *Right* operand recursively. Then **Solve** applies the operation on each pair in the Cartesian product of both sets *List1* and *List2* using two nested loops. In case the operation applies, the pair is collapsed using the **Merge** algorithm and the resulting component is added to *List3* the list of components matching the node. In case the list of components matching the node is not empty, then the node is said to be satisfied. As a precondition, the root of the subtree is an operator node with two leaf nodes representing operands.

Algorithm **Merge** in Figure 15 takes two composite components and merges them into one composite component. A composite component with elements of value Joker denotes a set of components since Joker can match any object. The Algorithm **Merge** goes over all elements of the composite components. In case both elements are Jokers it fills the resulting element with a Joker. In case one is a joker and the other is not, it fills the resulting component with the non-Joker element. In case both are not Jokers, and both are equal, it fills the resulting component with one of them. In case both are not Jokers and both are not equal, then a conflict is declared denoting that the intersection sets corresponding to the composite components is empty, and **Merge** returns a **null** object.

### 4.5.3. Operators Semantics

The `Compare()` function (Line 6, *Algorithm-1*) implements the semantics for the operators, which are listed and described in this section.

Binary positional operators are listed in Table 1. An operator takes as input two sets of components *Left* and *Right* and returns the set of element pairs $Res \subseteq Left \times Right$ that satisfy the operation semantics.

**Table 1.** *GUICop* positional operators and their semantics

| Operator | Purpose |
|---|---|
| $op_1$ above $op_2$ | Checks if the lower boundary of $op_1$ is above the upper boundary of $op_2$ |
| $op_1$ below $op_2$ | Checks if the upper boundary of $op_1$ is below the lower boundary of $op_2$ |
| $op_1$ leftto $op_2$ | Checks if the right boundary of $op_1$ is left to the left boundary of $op_2$ |
| $op_1$ righto $op_2$ | Checks if left boundary of $op_1$ is right to the right boundary of $op_2$ |
| $op_1$ contains $op_2$ | Checks if the left and right boundaries of $op_2$ are between the left and right boundaries of $op_1$, and the upper and lower boundaries of $op_2$ are between the upper and lower boundaries of $op_1$ |
| $op_1$ over $op_2$ | Checks if $op_1$ *and* $op_2$ are overlapping |
| $op_2$ leftaligned $op_2$ | Checks if the left boundary of $op_1$ is aligned with that of $op_2$ |
| $op_1$ rightaligned $op_2$ | Checks if the right boundary of $op_1$ is aligned with that of $op_2$ |
| $op_1$ topaligned $op_2$ | Checks if the top boundary of $op_1$ is aligned with that of $op_2$ |
| $op_1$ bottomaligned $op_2$ | Checks if the bottom boundary of $op_1$ is aligned with that of $op_2$ |
| $op_1$ smaller $op_2$ | Checks if the size (length or area) of $op_1$ is smaller than the size of $op_2$ |

The Boolean operator *not* takes as input a set of elements and returns its complement in the set of all captured elements. The conjunction *and* operator implements set intersection, and the disjunction *or* operator implements set union. The mutual exclusion *xor* operator takes two sets *Left* and *Right* and returns the elements in *Left* that are not in *Right* union those in *Right* and not in *Left*. The implication operator *implies* takes two sets *Left* and *Right* matching the left and right operands respectively. If *Left* is empty that means the left operand is false and any element satisfies the implication so the operator returns the universe. If *Left* is not empty, then the left operand is true and the elements that satisfy it must also satisfy the right operand for the implication to be true. Consequently, the intersection of *Left* with *Right* is returned..

The arithmetic operators +, -, *, /, the relational operators ==, <, >, !=, and the string operators *equals* and *concat* apply to the properties of the components declared in the variables section of the specification, e.g., r1.width + r2.width < r3.height. Additionally, in future work, *GUICop* will support temporal operators, string operators, and regular expression matching.

### 4.5.4. Convergence and Complexity Analysis

The algorithms presented in Figure 15 show how nodes in the AST graph are processed. The two nested loops in Algorithm **Solve** take $m_1 \cdot m_2$ steps to complete where $m_1$ and $m_2$ are the sizes of the component lists *List1* and *List2*, respectively. They invoke Algorithm **Merge** if needed. Algorithm **Merge** takes *v* steps to complete where *v* is the number of variables in the specification.

Algorithm **Solve** traverses the AST in a recursive manner with a base case at the leaf nodes. **Solve** therefore is an in-order depth first traversal of the AST tree. Both Algorithms **Solve** and **Merge** are therefore guaranteed to terminate.

Algorithm **Solve** is invoked *n* times where *n* is the number of nodes in the AST. In each invocation a Cartesian product between two sets is computed. Theoretically, the set matching a node at height $h$ from leave nodes can grow quadratic at every invocation resulting in a size of $m_{max}^{2^h}$, where $m_{max}$ is the maximum number of components matching a leaf node (a leaf nodes is at depth 0). Note though, the maximum size of a set of $v$-tuples is $m_{max}^{v}$ which is a polynomial bound. Consequently Solve is guaranteed to terminate within $n \min(m_{max}^{2^h}, m_{max}^{v})$ steps.

In practice, and due to (1) the use of Joker components to denote symbolic subsets and (2) the fact that practical constraints eventually eliminate non-matching tuples in the computation, all our experiments resulted in computation times well below both theoretical bounds above.

## 5. Case Studies and Experiments

In the context of *GUICop*, a *test case* is represented by: 1) a sequence of GUI *input* events (and possibly a test fixture) resulting in some GUI state; and 2) one or more *GUICop* specifications or test oracles that characterize the *expected* GUI state. Given a test case, the aim of the *GUICop Solver* is to verify that the expected GUI state is realized as a result of triggering its corresponding sequence of input events. In case the verification failed, i.e., the *Solver* failed to identify a set of rendered GUI objects that satisfy one or more of the specifications, *GUICop* produces an error message alarming the tester. Otherwise, *GUICop* produces a detailed report of the hierarchical objects that satisfy the specifications. The vision is to deploy *GUICop* as follows:

1) In its simplest configuration, *GUICop* would be applied by executing its *Driver* while no specifications are defined, i.e., by solely relying on the null-oracle. This is not an interesting configuration as it does not demonstrate the contribution of our work since existing techniques already provide similar capabilities.
2) In regression testing mode, *GUICop* would be applied by: a) executing a regression test suite, either manually or using some automated approach (e.g., *GUICop Driver*, GUITAR/GUI Ripping [24], capture/replay [29], or scripting); and b) defining a number of test oracles designed to make sure that some set of GUI scenarios of interest are still correctly rendered in the new version of the software. The scenarios could originate from user requirements or design use cases, or from previous bug fixes. The aim in the latter case is to verify that previously fixed bugs did not resurrect as a result of the code changes leading to the new version of the software [1][28]. As such, *GUICop* provides additional detailed oracles that complement the typically used null-oracle.

The focus in our experiments is to validate the ability of the *Solver* to detect non-crashing failures in real life applications, i.e., violations of *GUICop* user-defined test oracles. For that purpose, the researchers: 1) identified several Java/Swing subject programs with documented or injected non-crashing defects; 2) defined *GUICop* specifications characterizing the defects; 3) configured the *Code Weaver* to inject invocations to the *Solver* at locations that immediately follow the manifestation of failures; 4) manually executed test cases that induce the failures; and 5) examined the *Solver* output.

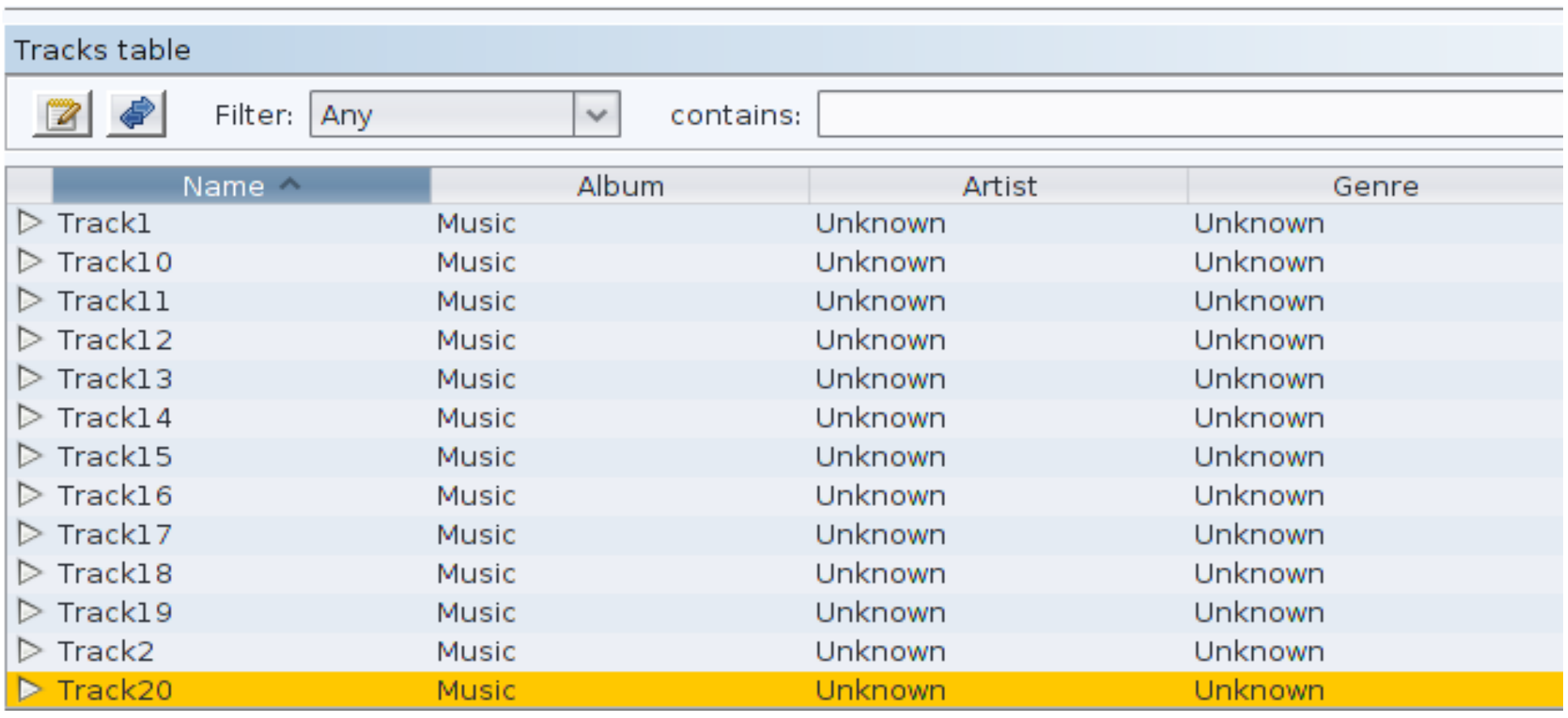

```
OrderedTracks = {
    variables {
        Textrect track_x;
        Textrect track_y;
    }
    properties {            // Defining properties is optional
        // by default X, Y, WIDTH and HEIGHT will describe the smallest rectangle
        // enclosing track_x and track_y, but in this case they have no role
    }
    constraints {
        (  ((track_x.text == 'Track2') and (track_y.text  == 'Track10'))  implies (track_x above track_y) );
    }
}
```

**Figure 16** – Faulty *Jajuk* display and corresponding *GUICop* specification.

The experiments involved four existing defects and four injected defects in four open source Java GUI applications that are based on the Swing GUI toolkit. The subject programs are: *Jajuk* (86K LOC), *Gason* (1.7K LOC), *JEdit* (301K LOC), and *TerPaint* (9.3K LOC). We also found an additional defect in TerPaint while experimenting with it. Please note that the *GUICop* toolset and case studies are available for download [5]. Presented next are the case studies, a discussion of the benefits of *GUICop* relative to other techniques, and a usability experiment.

### 5.1. *Case Study 1*: Jajuk

*Jajuk* (*www.jajuk.info*) is an advanced jukebox, a Java cross platform music organizer and player.

**Defect -** *GUICop* was applied on a real defect in *Jajuk* resulting in *"numbered tracks not being listed in order"*. When listing numbered tracks, *Jajuk* may fail to display them in the right order. For instance, Track1 is followed by Track10 instead of Track2 in the Tracks table shown in Figure 16.

**Oracle -** A *GUICop* specification named OrderedTracks was defined, also shown in Figure 16, which checks whether the tracks are listed in order. Specifically, OrderedTracks checks whether Track2 is displayed above Track10. OrderedTracks involves: 1) two Textrect variables track_x and track_y; 2) default properties; and 3) one constraint asserting that *if track_x contains "Track2" and track_y contains "Track10", then track_x must be rendered above track_y.*

***GUICop Solver* -** The *Code Weaver* was instructed to inject a call to the *Solver* at the appropriate location in the *Jajuk* code. A test fixture was set up and GUI events were manually triggered leading to the Tracks table to be rendered such that it contains more than eleven tracks with default names: Track1, Track2, …, Track10, Track11.

The *Solver* output was examined which logged the intermediate steps taken and a final statement indicating that OrderedTracks was violated. The end of Section 5.2 shows a sample *Solver* output.

### 5.2. *Case Study 2*: Gason

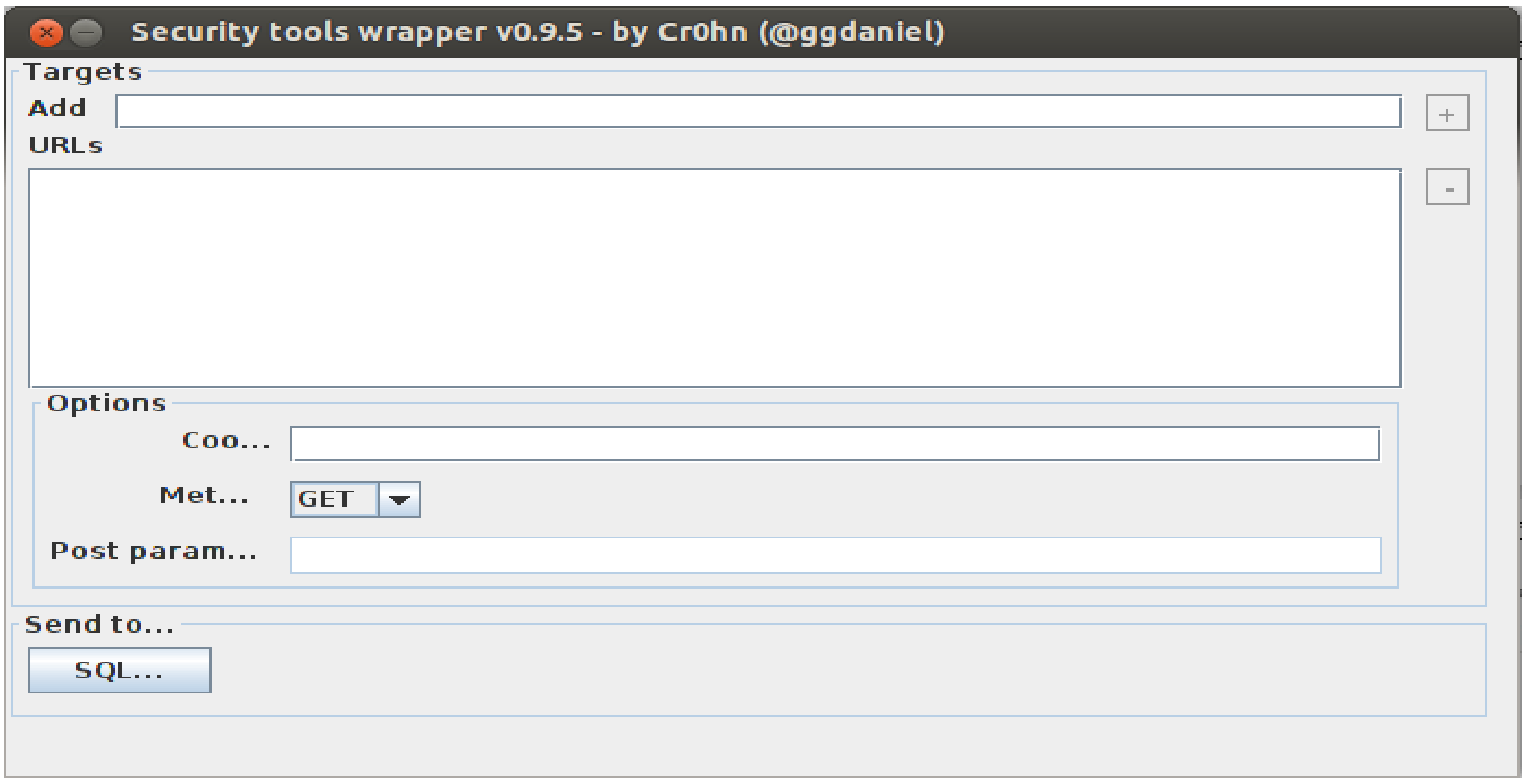

```
CroppedLabels = {
    variables {
        Textrect label1, label2, label3, label4;
    }
    properties { }
    constraints {          // Note: 'Send to…' is displayed properly
        (label1.text == 'Cookie');
        (label2.text == 'Method');
        (label3.text == 'Post parameter');
        (label4.text == 'SQLMap');
    }
}
```

**Figure 17** – Faulty *Gason* display and the CroppedLabels specification.

*Gason* is an open source plugin developed in Java to use *sqlmap* from *BurpSuite* (*portswigger.net/burp*), which is an integrated platform for performing security testing of web applications.

**Defect -** *GUICop* was applied on a real defect in *Gason* which causes some labels not to display words in full, i.e., the contents of the labels are in some cases cropped, as in the labels shown in Figure 17. The defect is reported and described in: "*Issue 3 - Contents of Listboxes cannot been seen completely - Available: https:// code.google.com/p/gason/issues/detail?id=3*".

**Oracle -** Figure 17 shows CroppedLabels, a *GUICop* specification that was defined to check whether four labels are cropped. It involves: 1) four Textrect variables label1, label2, label3, and label4; 2) default properties; and 3) one constraint asserting that label1 must show "Cookie", label2 must show "Method", label3 must show "Post parameter", and label4 must show "SQLMap". Note that the label showing "Send to..." is meant to be displayed as such.

***GUICop Solver* -** *GUICop* determined that the specification was violated once constraint "(label1.text == 'Cookie');" was not satisfied. The full Solver output is shown below:

```
0 CroppedLabels: .... [Textrect].label1.text strEquals 'Cookie'
0 CroppedLabels: .... [Textrect].label2.text strEquals 'Method'
0 CroppedLabels: .... [Textrect].label3.text strEquals 'Post parameter'
0 CroppedLabels: .... [Textrect].label4.text strEquals 'SQLMap'
0 CroppedLabels: all objects = [spec
{label4=[Textrect].label4, label3=[Textrect].label3, label2=[Textrect].label2, label1=[Textrect].label1}
 [[Textrect].label1.text strEquals 'Cookie'
, [Textrect].label2.text strEquals 'Method'
, [Textrect].label3.text strEquals 'Post parameter'
, [Textrect].label4.text strEquals 'SQLMap'
0 CroppedLabels: all textrects = [component 1 textrect (1,1,100,100,'Coo...'), component 2 textrect (1,1,100,100,'Met...'),
component 3 textrect (1,1,100,100,'Post param...'), component 4 textrect (1,1,100,100,'SQL...')]
0 CroppedLabels: Solving ... {
1   CroppedLabels: Starting match. {
1   CroppedLabels: matching variables {
2     Textrect:label4: Starting match. {
2     Textrect:label4: matchExecute
2     Textrect:label4: matchExecute done. found 4 objects
2     Textrect:label4: }. End match. Found 4 objects.
2     Textrect:label3: Starting match. {
2     Textrect:label3: matchExecute
2     Textrect:label3: matchExecute done. found 4 objects
2     Textrect:label3: }. End match. Found 4 objects.
2     Textrect:label2: Starting match. {
2     Textrect:label2: matchExecute
2     Textrect:label2: matchExecute done. found 4 objects
2     Textrect:label2: }. End match. Found 4 objects.
2     Textrect:label1: Starting match. {
2     Textrect:label1: matchExecute
2     Textrect:label1: matchExecute done. found 4 objects
2     Textrect:label1: }. End match. Found 4 objects.
1   CroppedLabels: }. done matching variables.
1   CroppedLabels: matching constraints {
1   CroppedLabels: Constraint 0 {
```

```
2    strEq: Starting match. {
2    strEq: matchExecute
3      label1.text: Starting match. {
4        Textrect:label1: Starting match. {
4        Textrect:label1: }. End match. Already computed. return cached.
3      label1.text:  i:0 s:'Coo...'
3      label1.text:  i:0 s:'Met...'
3      label1.text:  i:0 s:'Post param...'
3      label1.text:  i:0 s:'SQL...'
3      label1.text: }. End match. Found 4 objects.
3      string: Starting match. {
3      string: }. End match. Found 1 objects.
2    strEq: 'Coo...' str equals 'Cookie'
2    strEq: 'Met...' str equals 'Cookie'
2    strEq: 'Post param...' str equals 'Cookie'
2    strEq: 'SQL...' str equals 'Cookie'
2    strEq: done matchExecute. found 0 objects
2    strEq: }. End match. Found 0 objects.
1   CroppedLabels: }. Found 0 objects matching constraint 0.
1   CroppedLabels: }. done matching constraints. found 0 objects.
1   CroppedLabels: computing properties. {
1   CroppedLabels: } done computing properties.
1   CroppedLabels: }. End match. Found 0 objects.
0 CroppedLabels: }. Done solving.
0 CroppedLabels: Testing failed: specification not met!
```

### 5.3. JEdit

*JEdit* (www.jedit.org) is an open source Java text editor. *GUICop* specifications were written to guard against three defects in *JEdit*, of which one is real and two are injected. The real defect involves "Wrong Justification of Text". The first injected defect involves a "Missing HotKey Indicator", and the second involves a "Missing MenuItem Separator".

#### *5.3.1 Case Study 3*: ***Wrong Justification of Text***

**Defect -** *JEdit* supports over 160 character encodings for languages that are written left to right (LTR) and others that are written right to left (RTL) such as Arabic, Hebrew, and Urdu. However, when writing in an RTL language, the characters (wrongly) appear left justified.

**Oracle -** Figure 18 shows a *GUICop* specification that checks whether sentences of an RTL language appear right justified. `RightToLeft` declares `Editbox eb`, and two `Textrect t1` and `t2` meant to contain RTL sentences, and defines the following constraints:

1. *t1 and t2 should be contained within eb*
2. *t1 and t2 should be shorter than eb*
3. *t1 should be displayed above t2*
4. *if t1 contains the longer (RTL) sentence then the start of t1 should be to the left of that of t2*

5. *if t1 contains the shorter (RTL) sentence then the start of t1 should be to the right of that of t2*
6. *if the (RTL) sentences in t1 and t2 are of the same length then t1 and t2 should start at the same vertical location*

It should be noted that in RightToLeft, t1.text and t2.text appear to have the same width. However, depending on the font used, t1.width might be greater, less, or equal to t2.width. This is why the bottom constraint considers all three cases, and does not simply assume that t1 and t2 are of the same length.

***GUICop Solver*** **-** We configured the test fixture such that the two RTL sentences shown in Figure 18 are entered. Consequently, the *Solver* affirmed that the RightToLeft specification was violated.

Note that as part of future work the intention is to extend the *GUICop Specification Language*, which will then support the function "*boolean isRTL()*" in Textrect that returns true when the entered text is in an RTL language. When this is done, the bottom constraint in Figure 18 could instead be written as:

```
(    ( (t1.isRTL() == true) and ((t2.isRTL() == true) )
                    implies  (
                              (((t1.width > t2.width) implies (t1.x < t2.x)) and
                              ((t1.width < t2.width) implies (t1.x > t2.x))) and
                    ((t1.width == t2.width) implies (t1.x == t2.x)))          );
```

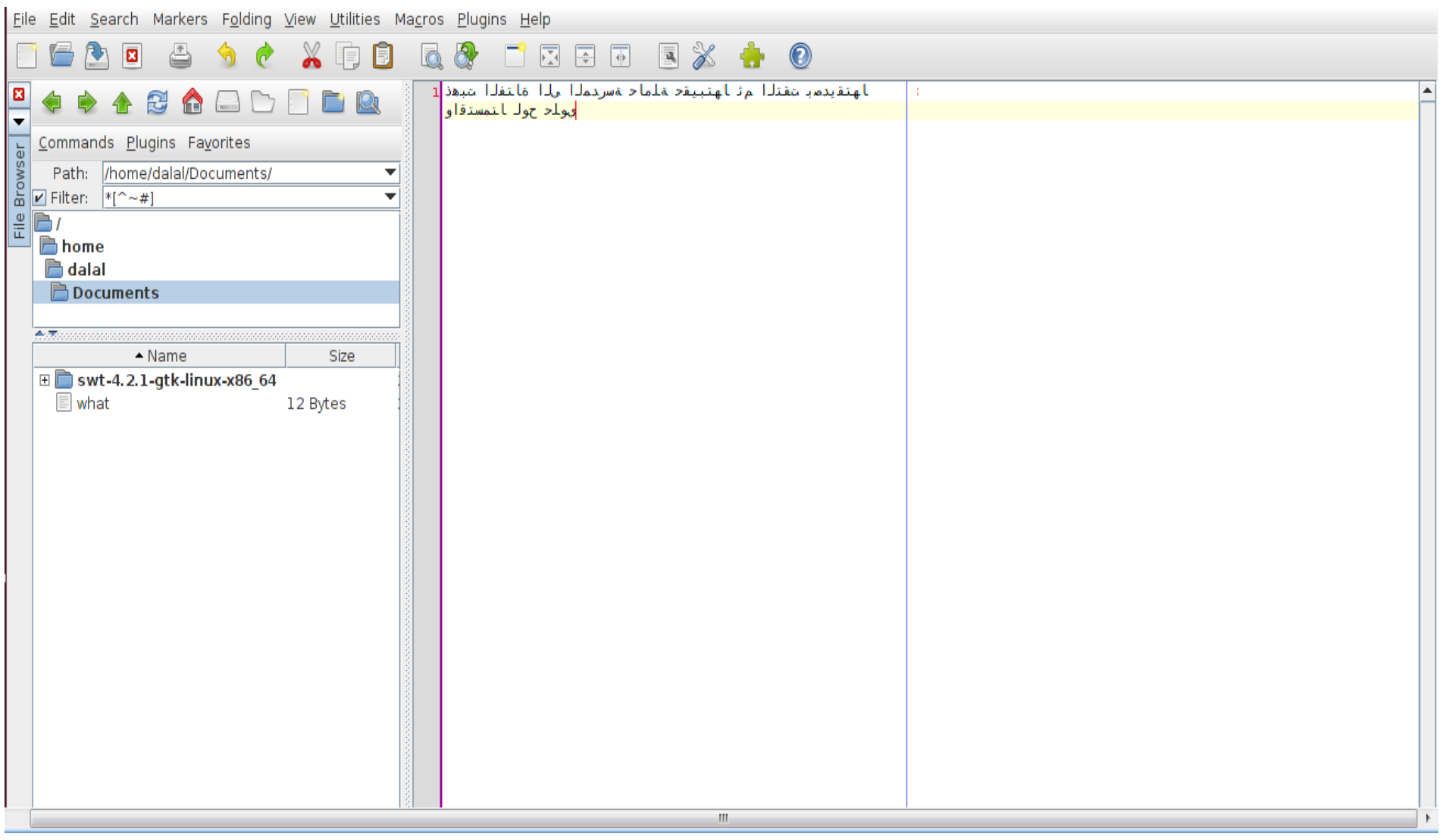

```
RightToLeft = {
      variables {
            Textrect t1, t2;
            Editbox eb;
      }
      properties {
            // by default X, Y, WIDTH and HEIGHT will describe
            // the smallest rectangle enclosing t1 and t2
      }
      constraints {
            // the concatenation of t1 and t2 yields a sentence stating that "the little girl went to
            // school carrying her bag then she met her friend with whom she shared a candy bar"
            (eb contains t1);
            (eb contains t2);
            (t1 above t2);
            (eb.width > t1.width);
            (eb.width > t2.width);

            (           ((t1.text equals 'حقيبتها ذهبت الفتاة الى المدرسة حاملة') and
                        (t2.text equals 'التقت بصديقتها و اقتسما لوح حلوى ثم'))
                        implies  (
                                    (((t1.width > t2.width) implies (t1.x < t2.x)) and
                                    ((t1.width < t2.width) implies (t1.x > t2.x))) and
                                    ((t1.width == t2.width) implies (t1.x == t2.x)))            );
      }
}
```

**Figure 18** – *JEdit*: right to left justification specification, which applies once the two RTL sentences above are entered. See translation in comments above.

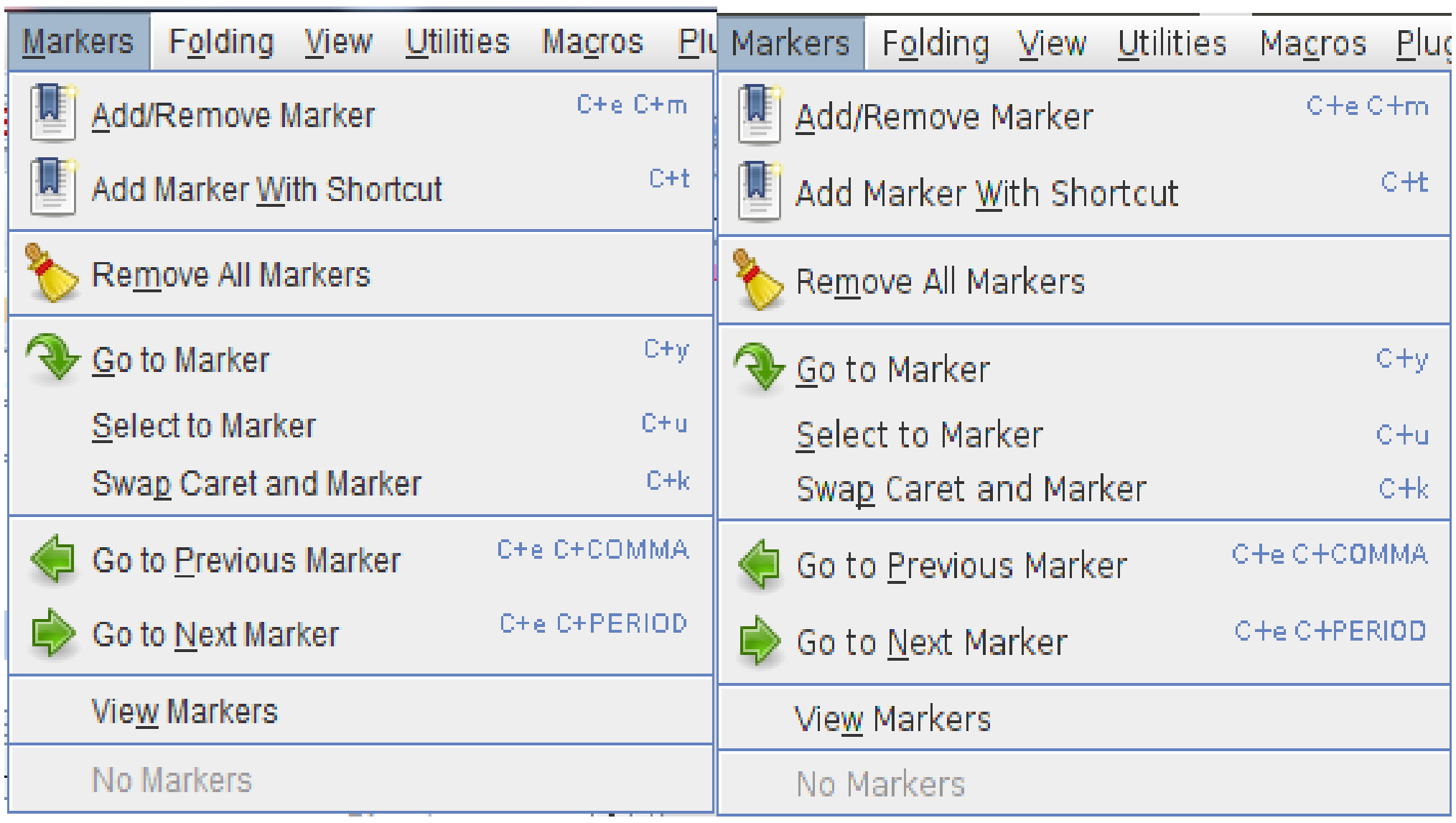

```
MainMenuLabel = {
        variables {
                ExtTextRect m;
                ExtRectangle hotkey;
        }
        properties {  }
        constraints {
                (m.text equals 'Markers');
                (m above hotkey);
                (hotkey.height < 3);          // rectangle with small height
                (m.x < (hotkey.x - 1));       // hotkey starts after text horizontally
                (hotkey.x2 < (m.x2 - 1));     // hotkey is shorter than text
                (hotkey.y > (m.y2 + 1));      // hotkey indicator is below text
                (hotkey.y < (m.y2 + 4));      // hotkey indicator is not way below text
        }
}

ExtRectangle = {
        variables { Rectangle r; }
        properties { x2 = (r.x + r.width); y2 = (r.y + r.height); }
        constraints { true; }
}
ExtTextRect = {
        variables { Textrect r; }
        properties { x2 = (r.x + r.width); y2 = (r.y + r.height); text = r.text; }
        constraints { true; }
}
```

**Figure 19** – *JEdit*: correct display, faulty display, and MainMenuLabel specification.

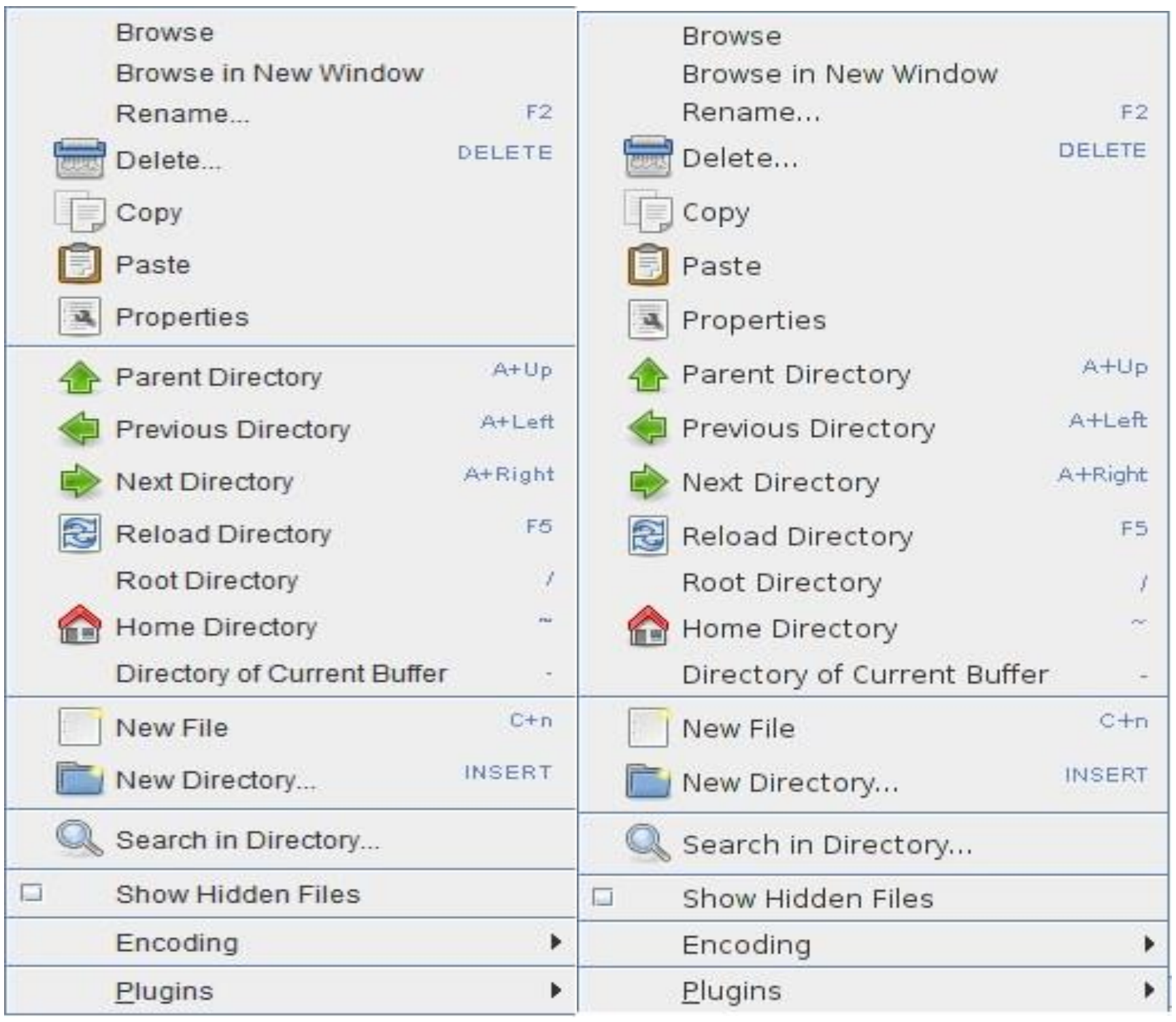

```
MenuWithSeparator = {
        variables {
                Textrect m1;
                Textrect m2;
                HLine sep;
        }
        properties { // default properties }
        constraints {
                ( ((m1.text equals 'Properties') and (m2.text equals 'Parent Directory'))
                                        implies ((m1 above sep) and (sep above m2) ) );
        }
}

HLine = {
        variables {  Line ln; }
        properties { // default properties }
        constraints {
                ((ln.y1 == ln.y2) and (ln.x1 < ln.x2));
        }
}
```

**Figure 20** – *JEdit*: correct menu items display, faulty display, and corresponding specifications.

*5.3.2 Case Study 4*: *Missing HotKey Indicator*

**Defect -** A defect was injected in *JEdit* by removing the hotkey indicator from the label of its *Markers* menu. Figure 19 shows the *Markers* menu with the hotkey indicator under the letter 'M'. It also shows it following the injection of the defect (note the missing hotkey from the label). Injecting the defect involved modifying the *jedit_en.props* properties file, and rebuilding *JEdit*.

**Oracle -** The specification MainMenuLabel, shown in Figure 19, was written to check against the absence of a hotkey in the label of the *Markers* menu. MainMenuLabel declares the variables: 1) m, an ExtTextRect representing the menu label; and 2) hotkey, an ExtRectangle representing the hotkey indicator. ExtTextRect is a specification, also shown in Figure 19, which defines a TextRect with three additional properties, x2, y2, and text. ExtRectangle is a specification which defines a Rectangle with two additional properties, x2 and y2.

The constraints section specifies the following: 1) *m should be displayed above hotkey*; 2) *hotkey is small in height*; 3) *the start of hotkey should be to the right of that of m*; 4) *hotkey is shorter than m*; and 5) *hotkey is mildly below m.*

***GUICop Solver* -** We applied *GUICop* on the mutated version of *JEdit* by first configuring the *Code Weaver* appropriately, then manually triggering the *Markers* menu to be displayed. The *Solver* indicated that the MainMenuLabel specification was violated.

*5.3.3 Case Study 5*: *Missing Menu Item Separator*

**Defect -** A menu item separator was programmatically removed from a context menu in *JEdit*. The left of Figure 20 shows a separator between menu items "Properties" and "Parent Directory". The right of the figure shows the separator missing. Injecting this defect involved modifying *BrowserCommandsMenu.java* by commenting out a call to the *addSeparator()* function.

**Oracle -** Figure 20 shows the MenuWithSeparator specification which was written to check whether the separator between the "Properties" and "Parent Directory" menu items is rendered. MenuWithSeparator declares the variables: 1) m1 and m2, two Textrect variables representing menu labels; and 2) sep, an HLine representing a separator. HLine is a specification, also shown in Figure 20, which defines a horizontal line.

The constraints section of MenuWithSeparator specifies the following: *if m1 contains "Properties" and m2 contains "*Parent Directory*", then m1 must be rendered above sep, which in turn should be rendered above m2.*

***GUICop Solver* -** We applied *GUICop* on the buggy version of *JEdit*, the Solver indicated that the constraint MenuWithSeparator was violated**.**

### 5.4. TerpPaint

*TerpPaint* (*sourceforge.net/projects/terppaint*) is an open source Java program providing capabilities similar to those of *Microsoft Paint*. It is part of the *TerpOffice* suite [35] developed at the University of Maryland by Atif Memon and his students. *TerpPaint* is ideal for evaluating *GUICop* as it is GUI intensive with non-trivial GUI capabilities, and unlike *Microsoft Paint*, its Java/Swing source code is readily accessible online (*www.cs.umd.edu/users/atif/TerpOffice*). We were not able to find any associated bug reports; alternatively, we identified one defect and injected two others to provide the basis for the case studies described next.

#### *5.4.1 Case Study 6: Failing to Resize of Main Canvas*

**Defect** – *TerpPaint* provides the user the ability to change the size of the main canvas by entering the desired width and height in pixels, inches, or centimeters. This is achieved using the "Attributes" dialog box accessible via the "Image->Attributes" menu item, shown in Figure 21. A defect was injected in the *okActionPerformed()* method implemented in *attributes.java*, which bypasses the *resizeImage()* call when the Pixels option is selected, thus resulting in the main canvas not being resized:

```
if (lastSelected == 1)              // Inches
      ((TerpPaint)this.getParent()).center.resizeImage((int)newPixW, (int)newPixH);
else if (lastSelected == 2)       // Centimeters
      ((TerpPaint)this.getParent()).center.resizeImage((int)newPixW, (int)newPixH);
else if (lastSelected == 2)       // Pixels – Injected Bug: the 2 was a 3
    ((TerpPaint)this.getParent()).center.resizeImage((int)currWidth, (int)currHeight);
```

**Oracle** – In order to test the functionality of changing the canvas size, a tester could enter the same pair of width and height (say 5 and 10) while selecting inches in one instance, then centimeters in another instance, and lastly pixels, to produce three

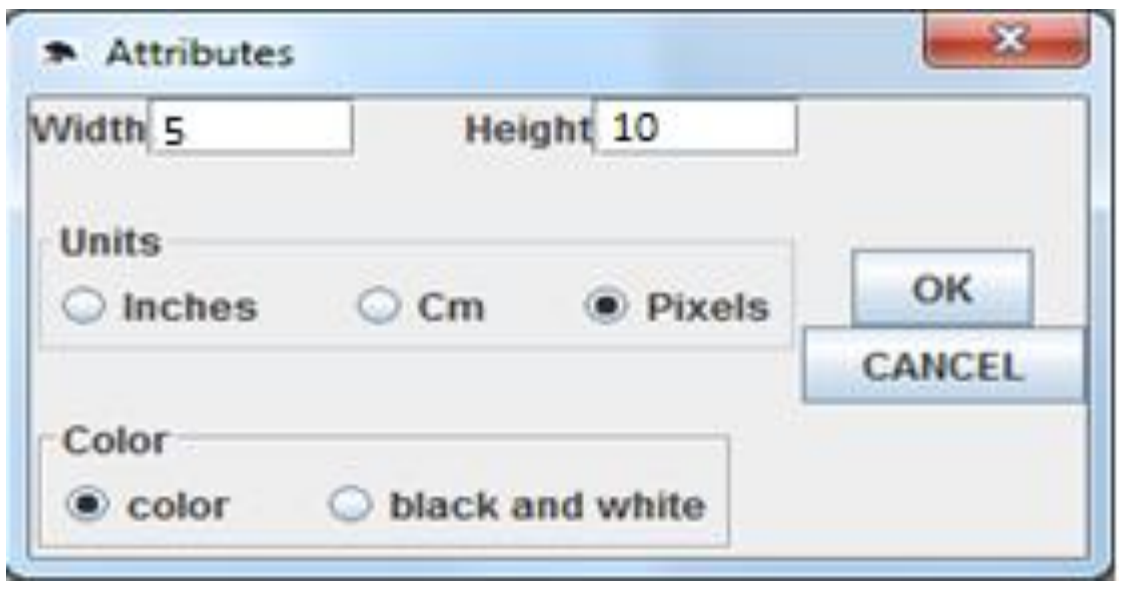

```
ResizedCanvas = {
        variables {
                Rectangle rInches, rCms, rPixels;
        }
        properties { // default properties }
        constraints {
                (rInches contains rCms);
                (rCms contains rPixels);
        }
}
```

**Figure 21** – *TerpPaint*: ResizedCanvas specification.

respective canvas sizes. The first size is expected to be the largest, followed by the second, then the third. However, due to the injected bug, the attempt to change the canvas while selecting pixels would not even produce a new canvas. Figure 21 shows the ResizedCanvas specification that checks whether there were three different canvas sizes observed, such that one canvas encloses another, which in turn encloses a third. This is a very simple specification involving three Rectangle objects called rInches, rCms, and rPixels, and two constraints specifying the following: *rInches contains rCms, and rCms contains rPixels.*

***GUICop Solver*** – The *Solver* was applied right after the third time canvas resizing was performed. Its output indicated that the ResizedCanvas specification was violated.

### *5.4.2 Case Study 7: Missing or Misspelled ToolTips*

**Defect** – The main window of *TerpPaint* contains 17 buttons with tooltip support. We injected two bugs in *TerpPaint.java*, one that removes the tooltip associated with the "Pencil" button, and another bug that misspells the text of the tooltip associated with the "Magnifier" button. Figure 22 shows snapshots of the misspelled tooltip and its corresponding correctly spelled form.

**Oracle** – A tester might want to check whether all 17 tooltips are properly supported. In support of that goal, Figure 22 shows CheckToolTips, a *GUICop* specification that validates whether all tooltips were properly displayed at the end of a testing session (whether manual or automated).

***GUICop Solver*** – We applied the *Solver* before and after injecting the bug, it first indicated that CheckToolTips was satisfied then it indicated that the specification was violated.

### ***5.4.3*** *Case Study 8: Misaligned OK and Cancel Buttons*

**Defect –** While using *TerpPaint* we realized that in all but one case, when a dialog box contains an 'OK/CANCEL' pair of buttons, the 'OK' button was precisely positioned to the left of the 'CANCEL' button. The exception was in the 'Attribute' dialog box in which they are positioned in a top down manner. Categorizing the latter as a bug is sensible. The top image in Figure 23 shows the buggy (current) behavior, and the bottom image shows the corrected behavior. The following describes the expected GUI behavior within a dialog box: *"In case there are both 'OK' and 'CANCEL' buttons, then 'OK' should be to the left of 'CANCEL' at the same height and without any overlap."*

**Oracle** – The CheckOKCancel specification, shown in Figure 23, checks whether the above expected behavior is satisfied. Specifically, the constraint in CheckOKCancel checks whether an 'OK' label and a 'CANCEL' label are both present; if either of them is not, the specification is satisfied. Otherwise, the specification is satisfied only if: 1) the top of the bounding boxes of both labels share the same y-coordinate (OK.Y == Cancel.Y); and 2) the bounding boxes do not overlap ((OK.X + OK.Width) < Cancel.X)).

***GUICop Solver* -** We applied the *Solver* before and after the bug was fixed, it first indicated that CheckOKCancel was violated then it indicated that the specification was satisfied.

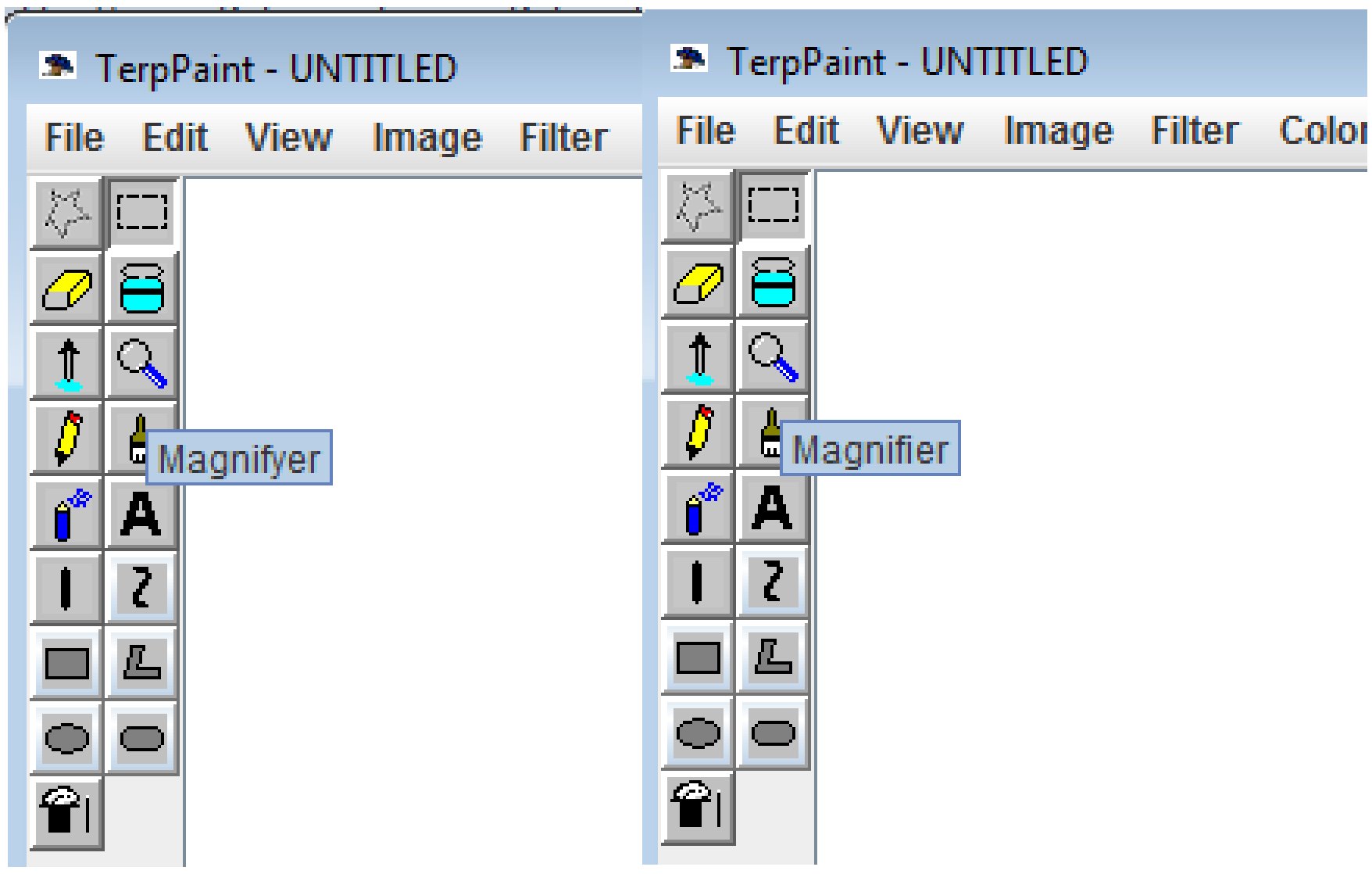

```
CheckToolTips = {
        variables {
                Textrect Line, Curve, RoundedRectangle, Rectangle, Polygon,
                        Ellipse, Select, Eraser, FillWithColor, PickColor, Magnifier,
                        Pencil, Brush, Airbrush, Text, FreeFormSelect, MagicWandSelect;

        }

        properties { }

        constraints {
                (Line.text equals 'Line');
                (Curve.text equals 'Curve');
                (RoundedRectangle.text equals 'Rounded Rectangle');
                (Rectangle.text equals 'Rectangle');
                (Polygon.text equals 'Polygon');
                (Ellipse.text equals 'Ellipse');
                (Select.text equals 'Select');
                (Eraser.text equals 'Eraser');
                (FillWithColor.text equals 'Fill With Color');
                (PickColor.text equals 'Pick Color');
                (Magnifier.text equals 'Magnifier');
                (Pencil.text equals 'Pencil');
                (Brush.text equals 'Brush');
                (Airbrush.text equals 'Airbrush');
                (Text.text equals 'Text');
                (FreeFormSelect.text equals 'Free-Form Select');
                (MagicWandSelect.text equals 'Magic Wand Select');
        }
}
```

**Figure 22** – *TerpPaint*: CheckToolTips specification.

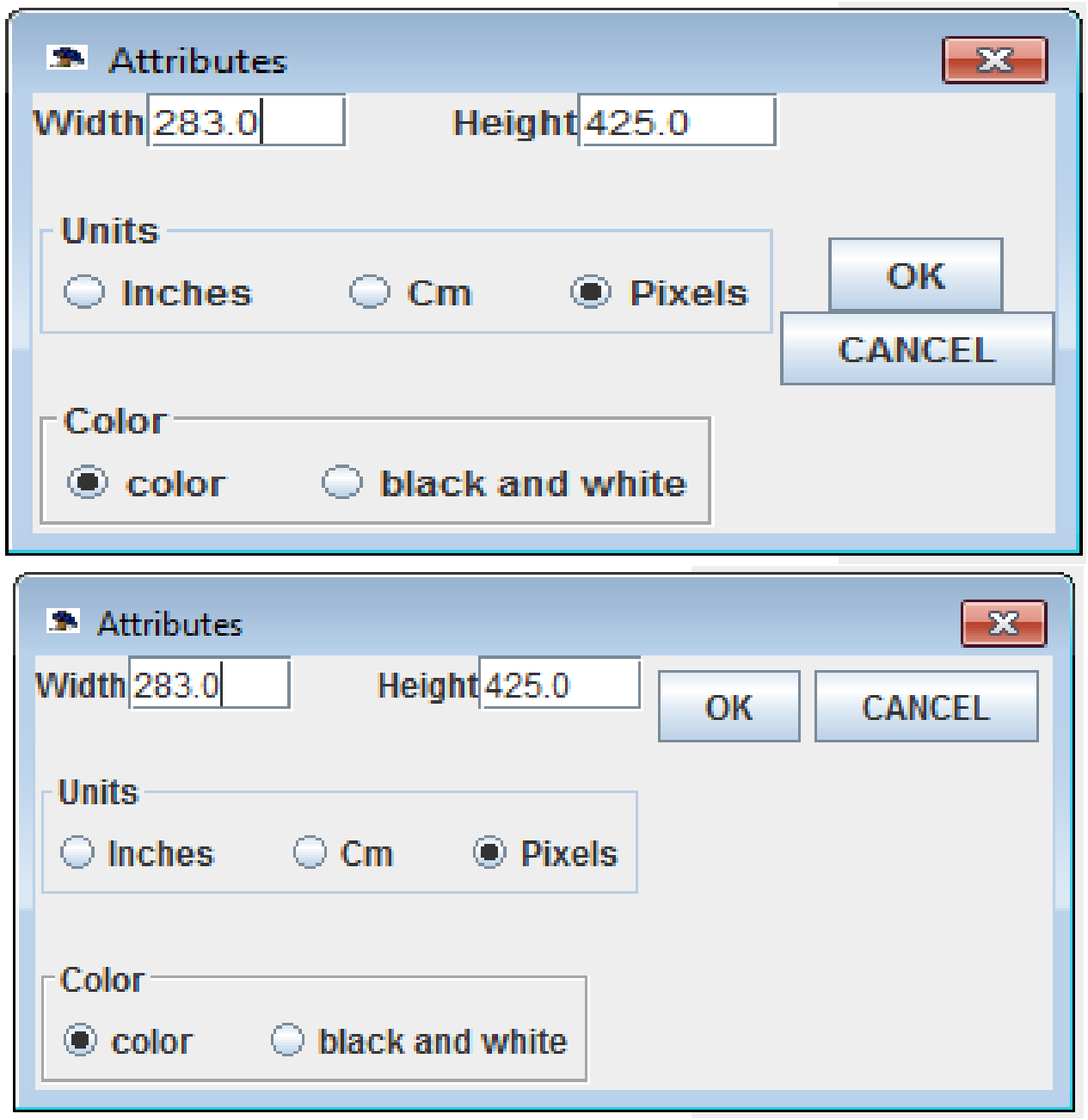

```
CheckOKCancel = {
        variables {
                Textrect OK, Cancel;
        }

         properties {  }

        constraints {
                ( ((OK.text equals 'OK') and (Cancel.text equals 'CANCEL')) implies
                                ((OK.Y == Cancel.Y) and ((OK.X + OK.Width) < Cancel.X)) );
        }
}
```

**Figure 23** – *TerpPaint*: CheckOKCancel specification.

While experimenting with *TerpPaint*, we encountered a failure that is exhibited as follows: 1) the resize dialog box is used to specify an excessive canvas size; 2) *TerpPaint* is closed; 3) *TerpPaint* is opened again; and 4) the observed failure is manifested by the toolbar pane being rendered twice, or by the main menu being rendered twice. We wrote a specification that characterizes either one of the manifestations of the defect and used it within *GUICop* to successfully detect the failing behavior.

### 5.5. Discussion

The presented case studies showcase the following:

1. *GUICop* is useful and applicable to real life applications.
2. The thoroughness of the *GUICop* oracles is valuable given that the null oracle cannot detect any of the defects in the case studies.
3. *GUICop* can be more beneficial than the GUI tree-based approaches given that in the case studies: a) names of GUI components were not known, e.g., the list items in *Jajuk (Case Study 1)*; and b) positional properties, such as *leftto*, cannot be checked directly or easily using the GUI tree relationships.
4. *GUICop* can be more beneficial than the image-based approaches, for example: a) in *Case Study 3*, RTL script detection may require *Optical Character Recognition* for the image, which is expensive and may not be reliable for non-Latin text; and b) in *Case Study 1*, checking that Track2 is above Track10 requires OCR, image segmentation and registration. Writing Sikuli scripts to specify such complex tasks requires image processing expertise.

### 5.6. Usability Experiments

In order to better assess the usability and user friendliness of *GUICop*, we conducted an experiment involving a number of users. We asked twelve students (elven undergraduates and one graduate) to write three *GUICop* specifications, one characterizing the defect in *Gason* (*Case Study 2*), one characterizing the third defect in *JEdit* (*Case Study 5*), and one introductory specification describing a robot-face shown below:

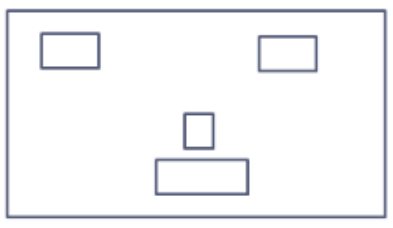

The students were unfamiliar with *GUICop* and the case studies. They received a 30 minutes presentation that included: 1) an overview of *GUICop* and its operators; and 2) an illustration of writing three specifications, specifically, HScrollbar (Figure 9), PushedRadiobutton (Section 4.1), OrderedTracks (*Case Study 1*). Table 2 summarizes the results of the experiment as follows: 1) the second column shows the minimum/maximum/average amount of time (in minutes) it took the students to write the given specification; 2) column three shows the number of submissions that successfully compiled; 3) column four shows the number of submissions that behaved correctly, i.e., given a GUI trace that satisfies the expected specification, did the *Solver* output a “pass”, alternatively, given a GUI trace that does not satisfy the expected specification, did the *Solver* output a “fail”; 4) column five shows the number of submissions that were partially correct; and 5) the last column shows the number of submissions that were incorrect.

**Table 2.** Results of the *GUICop* usability experiment

| Specification | Duration | | | # Compiled | # Correct | #Partially Correct | #Incorrect |
|---|---|---|---|---|---|---|---|
| | Min | Max | Avg | | | | |
| *Robot-Face* | 2 | 15 | 6.2 | 12 | 7 | 3 | 2 |
| *Case Study 2* | 2 | 13 | 6.1 | 12 | 11 | | 1 |
| *Case Study 5* | 2 | 13 | 5.4 | 12 | 8 | 3 | 1 |

The students’ submissions revealed the following interesting points:

1) The typical misunderstanding of the “implies” operator was exhibited in three submissions; i.e., $p \rightarrow q$ is actually *true* in case $p$ is *false*. We intend to address this issue by stressing this potential pitfall in the *GUICop* documentation and by supporting and *if-then-else* construct that might address the misunderstanding.
2) When evaluating composite expressions, *GUICop* constructs bounding rectangles around subexpressions, and uses their properties in the evaluation. This creates confusions with operators such as “above” and “leftto”. For example, considering constraint “((r1 above r2) above r3)” and the three rectangles below:

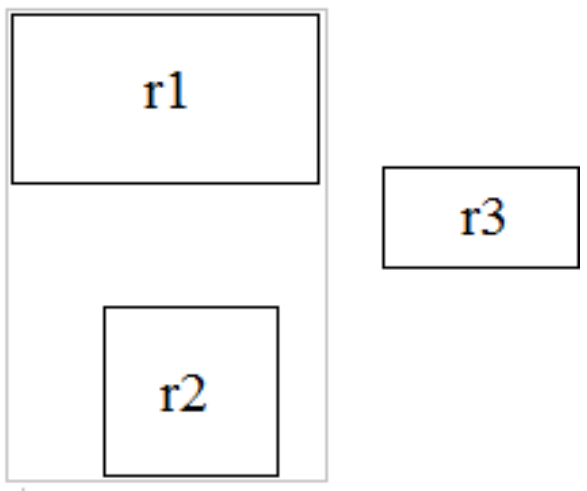

One might expect *GUICop* to determine that the constraint is not satisfied since r2 is not above r3. However, it determines that it is actually satisfied since the top of the bounding rectangle of “r1 above r2” (shown in grey) is above the top of r3. This confusion was exhibited in three submissions, which led us to reconsider the original behavior of *GUICop* and to modify it accordingly for the “above” operator and other similar ones such as “leftto”, “righto”, and “below”.

3) Several users assumed that one rendered shape may at most match one object within the specification. After further discussions, this exclusivity seemed intuitive for most users. Consequently, we adopted that as the default behavior in *GUICop* and extended support to a new keyword, namely “flexible”, that acts as a type modifier in variable declarations and indicates that the flexible variable may match objects matched by other variables as well.

Finally, the students were asked to provide feedback on the usability of the *GUICop Specification Language*. Below are some representative comments they wrote: *“Syntax familiar and easy to use”, “Intuitive language”, “What should be made very clear to the user is that the constraints should be satisfied for SOME x1, x2, x3 etc.”, “Was not sure how detailed the spec should be, e.g., was I supposed to also check whether the eyes were at the same level?”, “Had trouble writing all the constraints in a single one”*. In all, the students found the tool very easy to use, which could be partly attributed to the fact the GUI scenarios at hand were not very complex.

## 6. Threats to Validity

It could be argued that the functionality provided by *GUICop* does overlap with established existing approaches: 1) GUI Ripping [2][9][12] and HP WinRunner [6], due to its *Driver*; 2) GUI Modeling [3][18][19][20][21][22], due to its specification language; and 3) Sikuli [4][31], due to its verification capabilities. However, the main advantage of *GUICop* over the aforementioned approaches is its role as an accurate GUI oracle that is oblivious to GUI complicating factors such as changes in screen resolution or color scheme. It should also be noted that the above approaches could be extended with the checking capabilities of *GUICop*; particularly, 1) and 2).

We recognize the following threats to the validity of our approach:

1) *GUICop* relies on traces provided by the GUI libraries. So it correct modulo these libraries, and consequently it may be vulnerable when lower layers malfunction such as the graphics card, its device drivers, or its firmware.
2) *GUICop* requires learning a new language, which raises questions about its applicability. However, we believe that the proposed language is simple and intuitive as it exhibits syntactic similarities to popular languages such as Java and C++. The main elements not present in C++ and Java are new operators such as *leftto*, *above*, *contains*, and *smaller*. Furthermore, with a library of basic components readily available, users do not need to provide much detail to capture their specifications. Finally, the experiment we conducted in Section 5.6 demonstrates the usability of our approach.

3) In some cases users might write *GUICop* specifications that are detailed, and according to [24], detailed oracles pose two issues: 1) they limit opportunities for automation as they complicate the verification task; and 2) demand a higher cost of test maintenance. Regarding the first issue, *GUICop*'s verification is fully automated due to its *Solver*. As for the second issue, the *GUICop* specifications do not need to be updated for all types of GUI changes but only if the changes involved removing components that were used in the specifications or altering pre-specified layouts.
4) One might argue that detailed oracles are not necessary for GUI applications, and thus *GUICop* specification are not needed in practice. However, compared to the *null oracle*, *GUICop* specifications will most likely yield a lower rate of false negatives, given the additional detail they might provide. And compared to capture/replay, scripting, and computer vision techniques (e.g. Sikuli), *GUICop* is likely to yield a lower rate of false positives, given that variations in non-functional display parameters (e.g., screen resolution, color scheme) will not result in failures in the context of *GUICop*. It is sensible to say that the verification afforded by *GUICop* is not as loose as that of a *null oracle* and not as rigid as that of the many existing techniques, but again it all depends on the level of detail provided by the user. Furthermore, the study conducted by Xie and Memon [35] demonstrated that detailed (i.e., strong) oracles are more effective.
5) The benefits of the relatively high level of abstraction that *GUICop* operates at are not obvious. As previously stated, there are two main benefits to assert: 1) it allows *GUICop* to circumvent factors such as variations in screen resolution and other non-functional behaviors, which are problematic for most other techniques; 2) it allows for the reuse of specifications and for easily building more complex components owing to the provided *GUICop Specification Library*. Note though that the library might require some maintenance, e.g., if new basic components were to be supported by the library because they were missing or they needed to be re-implemented differently, their corresponding specifications will need to be defined and added. However, given the mature level of GUI libraries these days, such scenarios do not occur often. They typically occur during efforts for major rebranding or revamping of Operating Systems and GUI themes, e.g., the move from Windows XP to Windows 7. Considering the radio button example in Section 4.2, a change from an ellipse (or circle) to a rectangle will require an update of the specification. However, radio buttons are typically implemented using ellipses on most platforms, i.e., in practice such update is not likely to be needed.

## 7. Conclusions and Future Work

This paper presents *GUICop*, a new approach and a supporting toolset that checks whether the execution trace of a GUI program adheres to its user-defined specifications. *GUICop* specifications act as tester configurable oracles. The user-defined specifications aim at characterizing how GUI components are meant to be rendered, e.g., their layout, relative positioning, and visibility. *GUICop* is more practical than other existing techniques as it tolerates variations in screen resolution, color schemes, and line attributes such as style, thickness, and transparency.

As part of future work, the researchers will:

1) Explore the possibility of integrating *GUICop* within the GUITAR framework [24].
2) Extend the *GUICop* specifications to support: a) temporal operators in order to check event timings; b) more string operators; and c) regular expression matching; d) operators such as *intersects* and *occludes*; and e) properties such as font, style, line thickness, line type, and color.
3) Extend and simplify the *GUICop Specification Language* and provide a GUI tool that would facilitate writing *GUICop* specifications.
4) Allow for better reuse of existing specifications by extending the *GUICop Library* to include most widely used GUI components.
5) Allow for generic specifications to be checked in a global manner. For example, instead of requiring the user to configure the *Code Weaver* to inject `MainMenuLabel` checks at specific locations, checks would be performed every time a menu label is displayed.

**ACKNOWLEDGEMENT**
This research was supported in part by the Lebanese National Council for Scientific Research.